\documentclass[twocolumn]{aastex631}
\usepackage{graphicx}   
\usepackage{amsmath}	
\usepackage{hyperref}   
\usepackage[T1]{fontenc} 
\usepackage{mathptmx}
\usepackage{txfonts}
\usepackage{xspace}
\usepackage{gensymb} 
\usepackage{numprint}
\usepackage{booktabs}
\usepackage{threeparttable}

\begin{document}

\title{Prospects for Observing High-redshift Radio-loud Quasars in the SKA Era: Paving the Way for 21-cm Forest Observations}

\correspondingauthor{Xin Zhang}
\email{zhangxin@mail.neu.edu.cn}

\correspondingauthor{Yidong Xu}
\email{xuyd@nao.cas.cn}

\author[0009-0007-1168-0928]{Qi Niu}
\affiliation{Liaoning Key Laboratory of Cosmology and Astrophysics, College of Sciences, Northeastern University, Shenyang 110819, P. R. China}

\author[0000-0003-1962-2013]{Yichao Li}
\affiliation{Liaoning Key Laboratory of Cosmology and Astrophysics, College of Sciences, Northeastern University, Shenyang 110819, P. R. China}

\author[0000-0003-3224-4125]{Yidong Xu}
\affiliation{National Astronomical Observatories, Chinese Academy of Sciences, Beijing 100101, P. R. China}
\affiliation{Key Laboratory of Radio Astronomy and Technology, Chinese Academy of Sciences, Beijing 100101, P. R. China}

\author[0000-0003-4936-8247]{Hong Guo}
\affiliation{Shanghai Astronomical Observatory, Chinese Academy of Sciences, Shanghai 200030, P. R. China}

\author[0000-0002-6029-1933]{Xin Zhang}
\affiliation{Liaoning Key Laboratory of Cosmology and Astrophysics, College of Sciences, Northeastern University, Shenyang 110819, P. R. China}
\affiliation{National Frontiers Science Center for Industrial Intelligence and Systems Optimization, Northeastern University, Shenyang 110819, P. R. China}
\affiliation{MOE Key Laboratory of Data Analytics and Optimization for Smart Industry, Northeastern University, Shenyang 110819, P. R. China}

\begin{abstract}

The 21-cm forest is a sensitive probe for the early heating process and small-scale structures during the epoch of  reionization (EoR), to be realized with the upcoming Square Kilometre Array (SKA). Its detection relies on the availability of radio-bright background sources, among which the radio-loud quasars are very promising, but their abundance during the EoR is still poorly constrained due to limited observations. In this work, we use a physics-driven model to forecast future radio-loud quasar observations. We fit the parameters of the model using observational data of high-redshift quasars. Assuming Eddington accretion, the model yields an average lifetime of $t_{\rm q} \sim 10^{5.5}$yr for quasars at $z\sim6$, consistent with recent results obtained from quasar proximity zone observation. We show that if the radio-loud fraction of quasars evolves with redshift, it will significantly reduce the abundance of observable radio-loud quasars in the SKA era, making 21-cm forest studies challenging. With a constant radio-loud fraction, our model suggests that a one-year sky survey conducted with SKA-LOW has the capability to detect approximately 20 radio-loud quasars at $z\sim 9$, with sufficient sensitivity to resolve individual 21-cm forest lines.

\end{abstract}

\keywords{Quasars (1319) --- Radio loud quasars (1349) --- Reionization (1383) --- H I line emission (690)}


\section{Introduction} \label{sec:intro}

The origin and evolution of stars, galaxies, and supermassive black holes in the universe are among the great unsolved mysteries of cosmology. Unveiling these enigmas requires probing the first billion years of cosmic history. The exploration of the cosmic dawn and the epoch of reionization (EoR) primarily rely on the use of the 21-cm hyperfine structure spectral line of neutral hydrogen atoms. Detecting the 21-cm cosmological signal is extremely challenging, but once a breakthrough is achieved, its scientific significance is enormous \citep{2006PhR...433..181F,2010ARA&A..48..127M,2012RPPh...75h6901P,2020SCPMA..6370431X}.

Probing the 21-cm signals of hydrogen atoms in the early universe mainly relies on the cosmic microwave background as the background radio source, detecting the absorption signals from  the dark ages and the cosmic dawn, or the emission signals during the EoR when the gas is heated up by the early galaxies. Due to cosmic expansion, these 21-cm signals have been redshifted to low-frequency radio band, typically requiring low-frequency radio telescopes for observation. Currently, experiments to detect the cosmic dawn and EoR are in full swing, including experiments to detect the sky-averaged spectrum of the 21-cm signal during the cosmic dawn, such as the Experiment to Detect the Global EoR Signature (EDGES) \citep{2018Natur.555...67B}, the Shaped Antenna measurement of the background RAdio Spectrum (SARAS) \citep{2018ApJ...858...54S}, the Hongmeng Project \citep{2023ChJSS..43...43C}, the Mapper of the IGM Spin Temperature (MIST) \citep{2024MNRAS.530.4125M}, the Radio Experiment for the Analysis of Cosmic Hydrogen (REACH) \citep{2022NatAs...6..984D}, Probing Radio Intensity at high-Z from Marion ($\rm PRI^{Z}M$) \citep{2019JAI.....850004P}, the Large-aperture Experiment to Detect the Dark Age (LEDA) \citep{2018MNRAS.478.4193P}, Probing ReionizATion of the Universe using Signal from Hydrogen (PRATUTH) \citep{2023ExA....56..741S}, the Dark Ages Polarimeter Pathfinder (DAPPR)\footnote{\url{https://www.colorado.edu/project/dark-ages-polarimeter-pathfinder/}}, and the High-Z All-Sky Spectrum Experiment (HIGH-Z) \citep{Navros:2022qxa}. Additionally, there are experiments to detect the 21-cm power spectrum during the EoR, such as the Low Frequency Array (LOFAR) \citep{2017ApJ...838...65P}, the Murchison Widefield Array (MWA) \citep{2013PASA...30....7T}, the Precision Array for Probing the Epoch of Reionization (PAPER) \citep{2010AJ....139.1468P}, the Hydrogen Epoch of Reionization Array (HERA) \citep{2017PASP..129d5001D}, {the upgraded Giant Metrewave Radio Telescope (uGMRT) \citep{2017CSci..113..707G}, and the New Extension in Nançay Upgrading LOFAR (NenuFAR) \citep{2024A&A...681A..62M}.}
 
In particular, the Square Kilometre Array (SKA) \citep{2013arXiv1311.4288H} under construction will be the largest radio telescope to date. It has a large field of view, a broad band, and an ultra-high sensitivity, capable of taking tomographic images of ionizing intergalactic medium (IGM) \citep{2015aska.confE...1K}, which will reveal more detailed information of the EoR and make an important contribution to understanding the origin of cosmic structures.

In addition to these three measurement modes, another method of observing 21-cm signals is known as the 21-cm forest. If there are radio-bright point sources at the EoR, such as radio-loud quasars and radio afterglows of high-redshift gamma-ray bursts, neutral clumps along the line of sight would create individual absorption lines on the source spectra, forming forest-like features called the 21-cm forest (in analogy to the Lyman-$\rm \alpha$ forest) \citep{2002ApJ...577...22C,2002ApJ...579....1F,2006MNRAS.370.1867F,2009ApJ...704.1396X,2010SCPMA..53.1124X}. The observation of 21-cm forest is of great value for studying the properties of the IGM, as it is highly sensitive to the heating process caused by the formation of early galaxies \citep{2009ApJ...704.1396X,2011MNRAS.410.2025X,2013MNRAS.428.1755C}, and also for constraining the abundance of small-scale structures, so as to measure the properties of dark matter (DM) \citep{2014PhRvD..90h3003S,2020PhRvD.101d3516S,2021PhRvD.103b3521K}. Recently, it was found that the one-dimensional cross power spectrum measurement of the 21-cm forest can be used to help extract faint 21-cm signals, and can also further measure the properties of DM and the early galaxies simultaneously \citep{2023NatAs...7.1116S}.

The utilization of the 21-cm forest method is presently limited by three factors. Firstly, attaining the sensitivity necessary for detecting the 21-cm forest poses a challenge. Current radio telescopes such as LOFAR and MWA, lack the capability of conducting extensive detection deep into the EoR \citep{2013MNRAS.428.1755C,2024arXiv240510080C}. The anticipated completion of SKA in the near future will significantly mitigate this existing detection challenge \citep{2015aska.confE...6C,2015aska.confE...1K,2021MNRAS.506.5818S}. The vast effective aperture of SKA-LOW will empower us to detect the reionization process with unparalleled sensitivity. Secondly, there is a lack of an analytical model to connect physical parameters (such as DM mass and IGM temperature, etc.) with the 21 cm forest observational signals, making it difficult to carry out observational constraints on physical parameters (note that parameter inference based on large-scale simulations is extremely resource-intensive). To address this issue, we can consider constructing an analytical model. Alternatively, it is also possible to utilize methods based on deep learning. Recently, \cite{2024arXiv240714298S} have developed a set of deep learning-based data generation and likelihood-free parameter inference methods to simultaneously address the issues of the absence of an analytical model and small-scale non-Gaussianity. Thirdly, the 21-cm forest necessitates high-redshift radio point sources as backgrounds. Owing to observational limitations, our understanding of the abundance of high-redshift radio sources remains limited, constituting the foremost uncertainty in the application of the 21-cm forest.

Quasars are one of the brightest sources in the universe, with about 10\% of quasars exhibiting strong radio emissions \citep{2000AJ....119.1526S,2002AJ....124.2364I,2015ApJ...804..118B,2021A&A...656A.137G,2021ApJ...908..124L}. These radio-loud quasars serve as ideal background sources for 21-cm forest. With the continuous advancement of optical and near-infrared detection, the frontier of observing quasar redshift has reached 7.64 \citep{2021ApJ...907L...1W}. However, there are only {9} quasars observed with redshift exceeding 7 and 14 radio-loud quasars observed with redshift exceeding 6 \citep{2006ApJ...652..157M,2010AJ....139..906W,2015ApJ...804..118B,2021ApJ...909...80B,2021A&A...647L..11I,2021ApJ...908..124L,2022A&A...668A..27G,2023MNRAS.519.2060I,2023ApJS..265...29B,2023MNRAS.520.4609E,2024arXiv240707236B}. {Fortunately, the successful operation of the James Webb Space Telescope (JWST) has greatly enhanced our ability to detect high-redshift active galactic nuclei (AGNs). A number of high-redshift AGNs or AGN candidates have already been reported through observations with the JWST \citep[e.g.][]{2023A&A...677A.145U,2023ApJ...959...39H,2023ApJ...954L...4K,2023ApJ...953L..29L,2023ApJ...955L..24G,2024ApJ...966..229L}, making the search for AGNs at even higher redshifts possible.} During the deeper reionization epochs at higher redshifts, it is crucial to obtain theoretical estimates to determine whether there are sufficient radio-loud quasars available for 21-cm forest research.

To predict the abundance of high-redshift radio-loud quasars, \cite{2004ApJ...612..698H} (hereafter H04) developed a model driven by physical processes. By assuming a correspondence between supermassive black holes (SMBHs) and halos, as well as employing the Eddington limit for accretion, this model successfully derived the abundance of radio-loud quasars consistent with observational data. After nearly two decades of equipment upgrades and observational data accumulation, we have gained further understanding of some of the physical and statistical properties of high-redshift quasars \citep[e.g.][]{2011ApJS..196....2S,2012MNRAS.422..478R,2018ApJ...869..150M,2019ApJ...872L..29S,2021ApJ...923..262Y,2023ApJ...949L..42M,2023MNRAS.519.3027S,2024MNRAS.528.5692K}. Currently, some assumptions of the H04 model are challenged by observational data, requiring reasonable extensions and calibrations to fit the actual distribution of high-redshift quasars. This paper aims to update this model and make reasonable predictions for radio-loud quasar observations in the SKA era.

This paper is structured as follows. In Section\ \ref{sec:Method}, we describe the construction of the model and its calibration using high-redshift observational data. In Section\ \ref{sec:Result and Discuss}, we present predictions for future SKA-LOW observations of radio-loud quasars and discuss the impact of currently poorly constrained parameters on the results. In Section\ \ref{sec:Conclusion}, we give a summary of our findings and their implications. For this study, we presume the $\rm \Lambda$CDM model and use the cosmological parameters measured by Planck 2018 \citep{2020A&A...641A...6P}: $\Omega_m$ = 0.3111, $\Omega_{\Lambda}$ = 0.6889, $\Omega_b$ = 0.04897, $h$ = 0.6766, $\sigma_8$ = 0.8102, and $n_s$ = 0.9665.

\section{Method} \label{sec:Method}
In this section, we present the construction process of the model and discuss the rationality of some physical assumptions, eventually providing a calibrated model suitable for predicting high-redshift ($z\sim6$) quasar observations. The calibration of the model uses quasar UV luminosity functions data at $z\sim6$ \citep{2018ApJ...869..150M} and $z\sim6.8$ \citep{2023ApJ...949L..42M}, as well as the black hole masses and the DM halo masses they inhabit obtained from 49 quasars at $z\sim6$ \citep{2019ApJ...872L..29S}. 

\subsection{Halo Mass Function}
Quasars are powered by SMBHs. Therefore, we first need to estimate the abundance of SMBHs at the centers of galaxies. We assume that the SMBH population traces DM halos, and the distribution of DM halos follows the Sheth-Tormen form \citep{2002MNRAS.329...61S}, i.e. 
\begin{equation}
\frac{{\rm d}n}{{\rm d}M_{\rm halo}} = f(\sigma,z) \  \frac{\rho_{0}}{M_{\rm halo}} \frac{{\rm d}\ {\rm ln}\left( \sigma^{-1} \right)}{{\rm d}M_{\rm halo}}, \label{eq:0}
\end{equation}
where
\begin{equation}
f(\sigma,z) = 0.322\sqrt{\frac{3}{2\pi}}\frac{\delta_c}{\sigma} \left(1+{\left(\frac{4\sigma^2}{3{\delta_c}^2}\right)}^{0.3}\right) \exp{\left[- \frac{3{\delta_c}^2}{8\sigma^2}\right]}. \label{eq:1}
\end{equation}

Here $\delta_c = 1.686$ is the critical overdensity for spherical collapse, and $\rho_{0}$ is the background density. $\sigma$ is the r.m.s. of density fluctuations,
\begin{equation}
\sigma^2\left( M,z \right) = \sigma^2\left( R_{\rm sph},z \right) = \quad \int_0^\infty \frac{{\rm d}k}{2\pi^2} k^2 P\left(k,z\right) \left[ \frac{3j_1(kR_{\rm sph})}{kR_{\rm sph}} \right]^2,\label{eq:2}
\end{equation}
where $j_1(x)=({\rm sin} x-x\ {\rm cos}x)/x^2$, $R_{\rm sph}$ is the comoving radius related to mass $M$ by $M=4\pi \rho_{\rm m}{R^3_{\rm sph}}/3$.
The cosmological power spectrum $P(k,z)$ is calculated using the fitting form of \cite{1999ApJ...511....5E}. Since we are discussing quasars during the cosmic reionization when DM halos had not undergone large-scale mergers, this assumption is reasonable.

\subsection{Halo-SMBH Mass Relation}
By assuming that the radiation feedback and outflows of quasars determine the size of black holes, we can obtain the scaling relationship between black hole mass and DM halo mass \citep{1998A&A...331L...1S,1998MNRAS.300..817H,2003ApJ...595..614W}. The mass scaling relation is expressed as (Equation (4) in \cite{2003ApJ...595..614W}):
\begin{equation}
M_{\rm BH} = A\left(\frac{M_{\rm halo}}{1.5\times10^{12}M_{\odot}}\right)^{5/3} \left[\frac{\xi(z)}{\rm \Omega_m} \right]^{5/6} \left(1+z\right)^{5/2}M_{\odot} ,\label{eq:3}
\end{equation}
where $M_{\rm BH}$ is the black hole mass, $M_{\rm halo}$ is the DM halo mass, $A$ is the amplitude parameter, $\xi(z)$ is the dimensionless parameter related to redshift $z$, and $\xi(z) \simeq {\rm \Omega_m}$ for $z \geq 6$. 

{The scaling relation between $M_{\rm BH}$ and $M_{\rm halo}$ in Equation (\ref{eq:3}) is derived froma physically motivated model combining the $M_{\rm BH}-\sigma_{\rm h}$ relation ($M_{\rm BH} \propto {\sigma_{\rm h}}^{\beta}$, where $\sigma_{\rm h}$ is halo velocity dispersion) and the halo virialization ($M_{\rm halo} \propto \sigma_{\rm h}^3$). The radiation-driven model proposed by \cite{1998A&A...331L...1S} predicts $M_{\rm BH} \propto \sigma_{\rm h}^5$, while the momentum-driven model proposed by \cite{2003ApJ...596L..27K} predicts $M_{\rm BH} \propto \sigma_{\rm h}^4$. The results obtained from observations of the nearby universe lie between these two models \citep[e.g.][]{2000ApJ...539L..13G,2000ApJ...539L...9F,2009ApJ...698..198G,2013ApJ...764..184M,2013ARA&A..51..511K}, which are considered as evidence of the co-evolution of galaxies and SMBHs. However, this relation has not been thoroughly studied at high redshifts. To better predict the relation for high-redshift quasars, we use the SMBH and halo masses obtained from observational data of 49 quasars at $z\sim6$ \citep{2019ApJ...872L..29S}. We tried to treat the first power-law index in Equation (\ref{eq:3}) as a free parameter, and found that the 49 data points favor a notably low power-law index of $\sim$ 0.22. Such a low index shows large discrepancy from low-redshift observations, and it could hardly be explained by physically-motivated models. Considering the large intrinsic scatter and the large error bars of the limited sample, we attribute this discrepancy to observational systematics and incompleteness at high redshifts, which prevent us from obtaining a reliable index estimate from these data. Due to the limited understanding of the evolution of SMBHs at high redshifts, in this study we follow the choice of H04 and adopt a fixed index of 5/3.}

\begin{figure}
\includegraphics[width=0.47\textwidth]{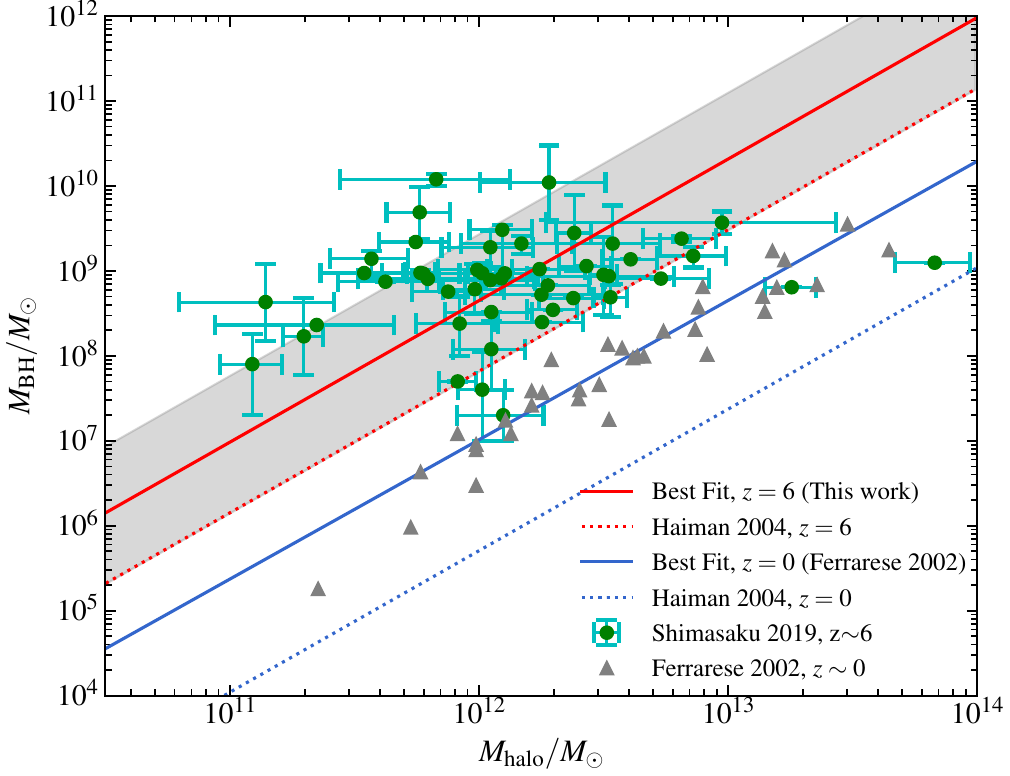}
\caption{$M_{\rm halo}-M_{\rm BH}$ relation. The red lines represent the fitting results at $z=6$, and the blue lines represent results for $z=0$. Solid lines represent observational fitting, while dashed lines represent the mass relation used by H04. The shaded area represents the intrinsic scatter $\sigma_{\text{in}}$. The gray triangular data points are from \cite{2002ApJ...578...90F}, and the green data points are from \cite{2019ApJ...872L..29S}.
\label{fig:MMrelation}}
\end{figure}

{We note that the $M_{\rm BH}-\sigma_{\rm bulge}$ relation measured in the nearby universe exhibits intrinsic scatter \citep[e.g.][]{2013ARA&A..51..511K}, which will naturally propagates to the $M_{\rm halo}-M_{\rm BH}$ relation we are fitting. Therefore, we assume that the high-redshift data contain intrinsic scatter following a Gaussian distribution $\mathcal{N}(0,\sigma_{\text{in}})$ in the logarithmic space, and construct the likelihood function as
\begin{alignat}{2}
 &\mathcal{L}(A,\sigma_{\text{in}}) = \prod_{i=1}^{N} \frac{1}{\sqrt{2\pi (\sigma_{m,i}^2+\sigma_{\text{in}}^2)}} \nonumber\\
 &\times \exp\left( -\frac{\left(\ {\rm log}\left(M_{{\rm BH},i}\right) - {\rm log}\left(M_{{\rm model},i}(A)\right)\ \right)^2}{2 (\sigma_{m,i}^2+\sigma_{\text{in}}^2)} \right) ,\label{eq:likelihood}
\end{alignat}
where $A$ is the amplitude parameter defined in Equation~(\ref{eq:3}), $\sigma_{\text{in}}$ is the intrinsic scatter, $\sigma_{m,i}$ is the measurement error of each data point, $M_{{\rm BH},i}$ is the black hole mass of each data point, and $M_{{\rm model},i}$ is the black hole mass obtained from Equation (\ref{eq:3}). We obtain the best-fit values of the amplitude parameter $A$ and the intrinsic scatter $\sigma_{\text{in}}$ using the maximum likelihood estimation method, and the results are shown in Figure \ref{fig:MMrelation} as the red line with the shaded area being the $1\sigma$ range (${\rm log}(A)=6.8\pm 0.81$), the reduced $\chi^2\simeq1.02$. We found that the best-fit amplitude parameter $A=10^{6.8}$ is approximately 7 times the amplitude used by H04. As the intrinsic scatter is well-characterized by a normal distribution, the fitting result of Equation (\ref{eq:3}) represents the theoretical mean. For simplicity, we exclusively use Equation (\ref{eq:3}) with the fitted amplitude in subsequent calculations, without accounting for intrinsic scatter.
}

We note that H04 used quasar luminosity function data at $z\gtrsim3$ to fit the amplitude parameter, with the average quasar lifetime fixed at $2\times10^7$ years. However, the results obtained did not match the observations of the local universe and those at $z=6$ (see Figure \ref{fig:MMrelation}). Note that recent studies on quasar lifetimes suggest shorter lifetimes \citep{2017ApJ...840...24E,2021ApJ...917...38E,2021ApJ...921...88M}, and in our analysis, the average quasar lifetime is treated as a variable and is constrained by high-redshift luminosity function data (see Section \ref{subsec:Duty Cycle}). 

\subsection{Accretion Rate and Optical Luminosity} \label{subsec:Accretion}
We assume that the average Eddington ratio $\lambda_{\rm Edd}$ for high-redshift quasars is 1, indicating Eddington accretion. Although the average accretion rate of quasars in the local universe is much lower than Eddington accretion, recent high-redshift observations suggest that the average Eddington ratio of early quasars is close to 1 \citep{2021ApJ...923..262Y}. Additionally, we note that to explain the formation of SMBHs represented by the observed high-redshift quasars, some studies have adopted models of sustained mild super-Eddington accretion \citep[e.g.][]{2015MNRAS.454.3771P,2023ApJ...950...85L} or chaotic accretion \citep[e.g.][]{2021MNRAS.501.4289Z}. The results of \cite{2023arXiv231208422L} indicate that super-Eddington accretion can also persist for a long time. Based on the above observation and simulation results, we believe our assumption is reasonable.

Once we determine the Eddington ratio, we can calculate the optical luminosity of quasars. Assuming a quasar has an average Eddington ratio of $\lambda_{\rm Edd}$ over its lifetime, we can get the bolometric luminosity as
\begin{equation}
L_{\rm bol} = \lambda_{\rm Edd} L_{\rm Edd}, \label{eq:5}
\end{equation}
where $L_{\rm Edd}$ is the Eddington luminosity. As the discussion above, we adopt $\lambda_{\rm Edd}=1$ for most of our calculations. We will also discuss the case with $\lambda_{\rm Edd}=1.2$.

For the quasar bolometric luminosity corrections, we use the results from \cite{2012MNRAS.422..478R} (their Equations (11) and (13)):
\begin{equation}
{\rm log}\left(L_{\rm bol}\right) = {\rm log}\left(0.75\right) + 4.89 + 0.91\times {\rm log}\left(5100L_{5100}\right) \label{eq:6} .
\end{equation}
We adopt the average spectral energy distribution {(SED)} from \cite{2011ApJS..196....2S} to calculate the luminosity relationship between different optical bands: 
{\begin{equation}
\lambda F_{\lambda}\left(1450 \AA\right) \simeq 1.86\ \lambda F_{\lambda}\left(5100 \AA\right), \label{eq:SED1450}
\end{equation}
\begin{equation}
\lambda F_{\lambda}\left(2500 \AA\right) \simeq 1.76\ \lambda F_{\lambda}\left(5100 \AA\right), \label{eq:SED2500}
\end{equation}
\begin{equation}
\lambda F_{\lambda}\left(4400 \AA\right) \simeq 1.11\ \lambda F_{\lambda}\left(5100 \AA\right), \label{eq:SED5100}
\end{equation}
where $\lambda F_{\lambda}(\lambda)$ is the rest-frame spectral flux at wavelength $\lambda$.}

\begin{figure}
\includegraphics[width=0.47\textwidth]{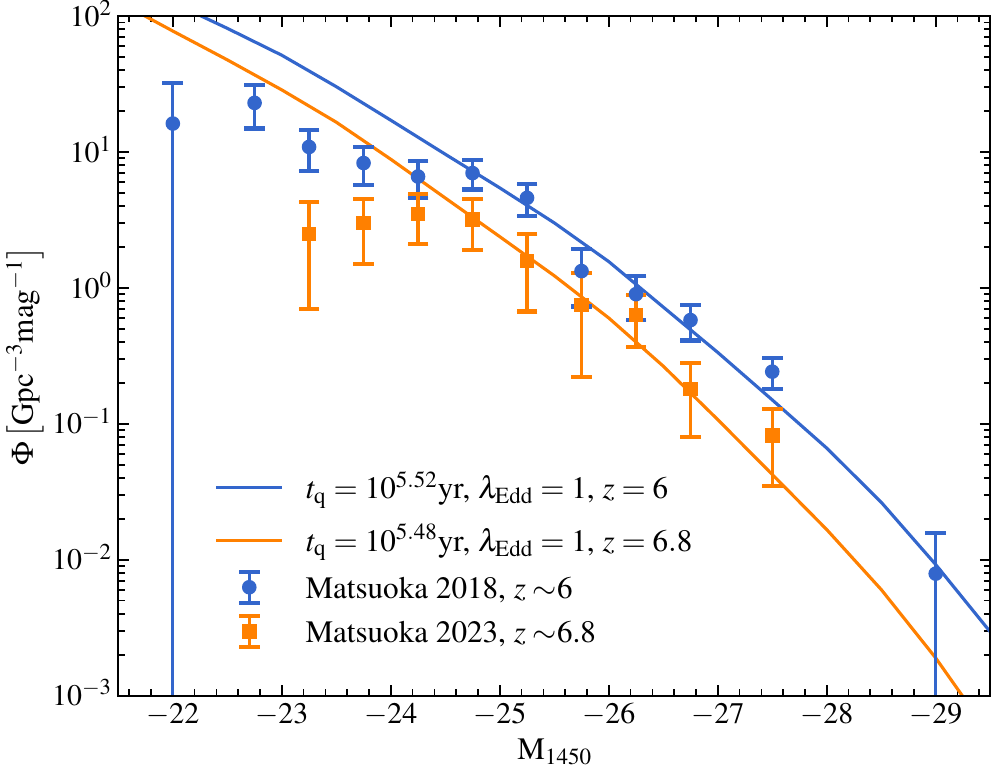}
\caption{UV luminosity function fitting results. The average Eddington ratio of quasars is fixed at 1. The blue line represents $z=6$, and the orange line represents $z=6.8$. Solid lines represent model best fits. Note that we ignored the first two points on the faint end during fitting due to observational incompleteness. The data points are provided by \cite{2018ApJ...869..150M} and \cite{2023ApJ...949L..42M}.
\label{fig:QLFfit}}
\end{figure}

\subsection{Duty Cycle of Quasar Phase}\label{subsec:Duty Cycle}
Through the halo mass function and the halo--SMBH mass relation, we have obtained the {black hole mass function.} 
However, not all black holes are in the quasar phase at the same time. We need to estimate the proportion of black holes that are in the quasar phase, which is the duty cycle. We estimate the duty cycle as
\begin{equation}
D_{\rm q} = \frac{t_{\rm q}}{t_{\rm H}(z)}, \label{eq:Dq}
\end{equation} 
where $t_{\rm q}$ is the average lifetime of quasars, and $t_{\rm H}(z)$ is the age of the universe at redshift $z$.

{By combining the black hole mass function with Equations (\ref{eq:5})--(\ref{eq:SED1450}) and (\ref{eq:Dq}), we can calculate the quasar luminosity function (QLF, the comoving number density of quasars per unit of magnitude) as
\begin{equation}
\Phi \left(M_{1450},z\right) = \frac{{\rm d}n_{\rm q}}{{\rm d}M_{1450}} = D_{\rm q}\frac{{\rm d}n_{\rm BH}}{{\rm d}M_{\rm BH}} \frac{{\rm d}M_{\rm BH}}{{\rm d}M_{1450}},
\end{equation} 
where $n_{\rm q}$ is the comoving number density of quasars, ${\rm d}n_{\rm BH}/{\rm d}M_{\rm BH}$ is the black hole mass function, and $M_{1450}$ is  the absolute magnitude at $1450\ \AA$,
\begin{equation}
M_{1450} = -2.5 {\rm log}\left(\frac{L_{\nu}\left(\nu(1450 \AA)\right)}{4\pi d^2 3631{\rm \ Jy}} \right),
\end{equation}
where $d=10\ {\rm pc}$ is expressed in cm.}

{In the calculation of QLF, $\lambda_{\rm Edd}$ and $t_{\rm q}$ are two parameters that are not strictly constrained by observations. Since both jointly determine the shape of QLF, they are strongly correlated parameters when fitting QLF to observations. Based on the discussion of the accretion rate in Section \ref{subsec:Accretion}, we obtained the best fit to the observed luminosity function data by varying the average lifetime $t_{\rm q}$ of the quasars until $\chi^2$ is minimized, using a fixed average Eddington ratio $\lambda_{\rm Edd} = 1$ for early quasars.} We employed two sets of high-redshift observational data, obtained from observations of $z\sim6$ \citep{2018ApJ...869..150M} and $z\sim6.8$ \citep{2023ApJ...949L..42M}. Our best-fit results are shown in Figure \ref{fig:QLFfit}. {The reduced $\chi^2$ obtained from the fits at $z=6$ and $z=6.8$ are 1.49 and 0.79, respectively.} We take the logarithmic mean value $ 10^{5.5}$ yr in subsequent calculations. Note that both sets of data, especially the $z\sim6.8$ data, are affected by incomplete observation coverage at the faint end, so we did not use the data points at the faint end with $M_{1450}>-24$ in the fitting process. 

\begin{figure}
\includegraphics[width=0.47\textwidth]{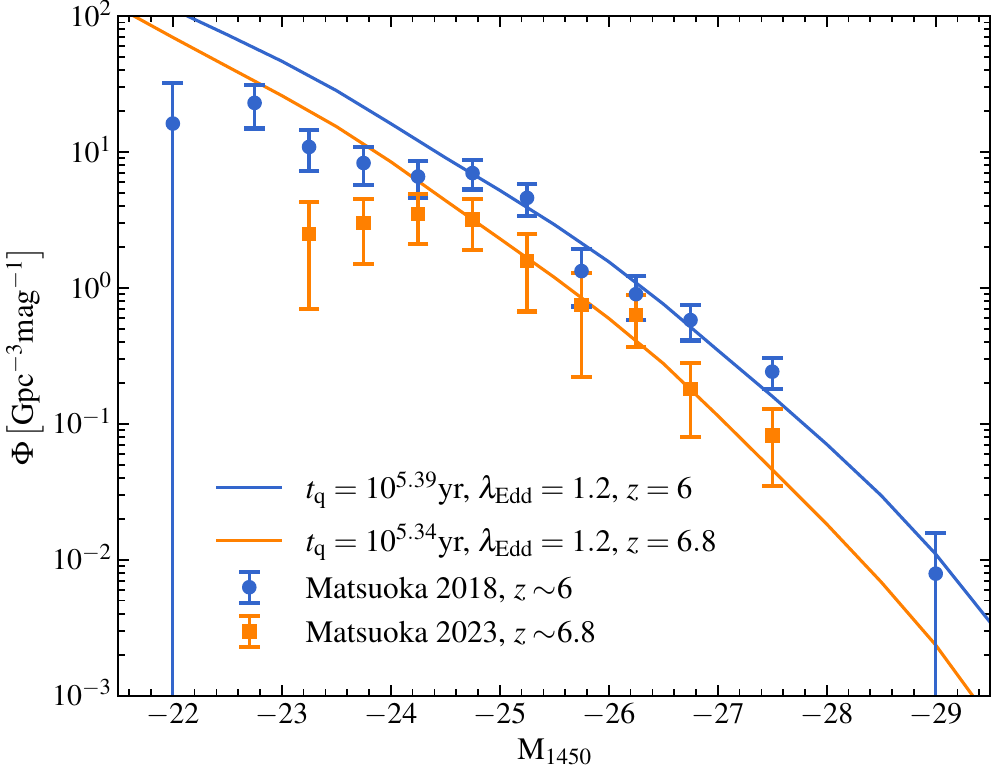}
\caption{Similar to Figure \ref{fig:QLFfit}, but the average Eddington ratio of quasars is fixed at 1.2.
\label{fig:QLFfit1}}
\end{figure}

As a possible scenario, we also calculated the fit with an average Eddington ratio of $\lambda_{\rm Edd}=1.2$, which is shown in Figure \ref{fig:QLFfit1}. {The reduced $\chi^2$ obtained from the fits at $z=6$ and $z=6.8$ are 1.26 and 0.68, respectively. From the results, it seems that the fits with a higher $\lambda_{\rm Edd}$ is better, but we must point out that this result could still be influenced by the incomplete observations of high-redshift QLF. The incompleteness effect can be divided into two parts. First, the sensitivity limitation, where there is incompleteness in the data at the faint end of QLF, and we mitigate this effect by discarding the data points with $M_{1450}>-24$. Second, the obscuration by the AGN dust torus may affect the completeness of optical observations, meaning that some AGNs cannot be effectively detected in the optical band.}

We note that \cite{2018MNRAS.478.5564B} used a constant duty cycle of $D_{\rm q} = 0.02$ to fit the observed optical luminosity function. However, \cite{1998ApJ...503..505H} and \cite{1998MNRAS.300..817H} pointed out a connection between the duty cycle and Equation (\ref{eq:3}). If the amplitude factor $A$ in Equation (\ref{eq:3}) is changed, the resulting duty cycle will also change. Since the amplitude factor $A$ we use is different from that of \cite{2018MNRAS.478.5564B}, the resulting duty cycle is also different. Note that our amplitude factor $A$ and duty cycle $D_{\rm q}$ are fitted from observational data at $z \sim 6$, so the resulting duty cycle is only applicable to high redshifts.

\subsection{Radio-loud Fraction}
We assume that the radio-loud fraction (RLF) of quasars at high redshift is 10\%. This assumption is consistent with observational studies of the RLF up to redshift of $z\sim6$ \citep{2000AJ....119.1526S,2002AJ....124.2364I,2015ApJ...804..118B,2021A&A...656A.137G,2021ApJ...908..124L}. We also note that \cite{2007ApJ...656..680J} proposed a form for the evolution of the RLF with redshift and luminosity,
\begin{equation}
{\rm log}\left(\frac{\rm RLF}{1-\rm RLF}\right) = b_0+b_z{\rm log}\left(1+z\right)+b_M \left(M_{2500}+26\right),
\end{equation}  
where $M_{2500}$ represents the optical magnitude at $2500\ \AA$. $b_0$, $b_z$ and $b_M$ are free parameters, with best-fit values of $-0.132$, $-2.052$, and $-0.183$, respectively. Since the recent study on the RLF of $z\geq6$ quasars does not rule out the possibility of this form of evolution \citep{2024MNRAS.528.5692K}, we also calculated its impact on the results (see Section \ref{subsec:RLFimpact}).

\begin{figure}
\includegraphics[width=0.47\textwidth]{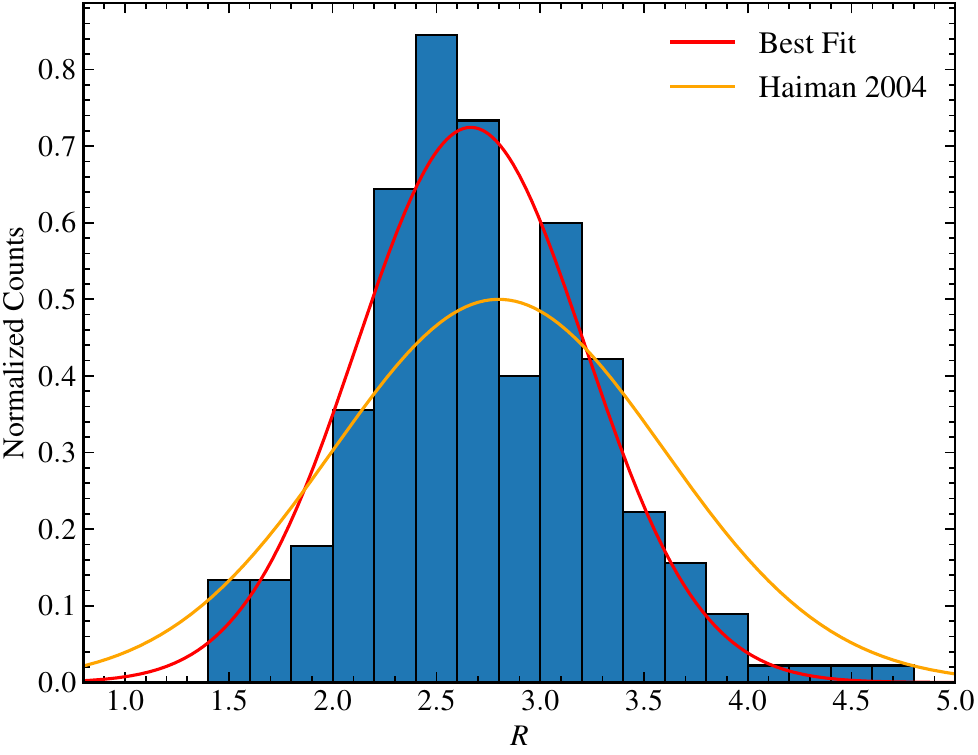}
\caption{The fitting results of the radio-loudness distribution. The radio-loudness data used are from \cite{2022MNRAS.513.4673B}. The red line represents our best fit to the data, and the orange line represents the radio-loudness distribution used in H04.
\label{fig:Radioloudness}}
\end{figure}

\subsection{Radio-loudness Distribution}
\begin{figure}
\includegraphics[width=0.47\textwidth]{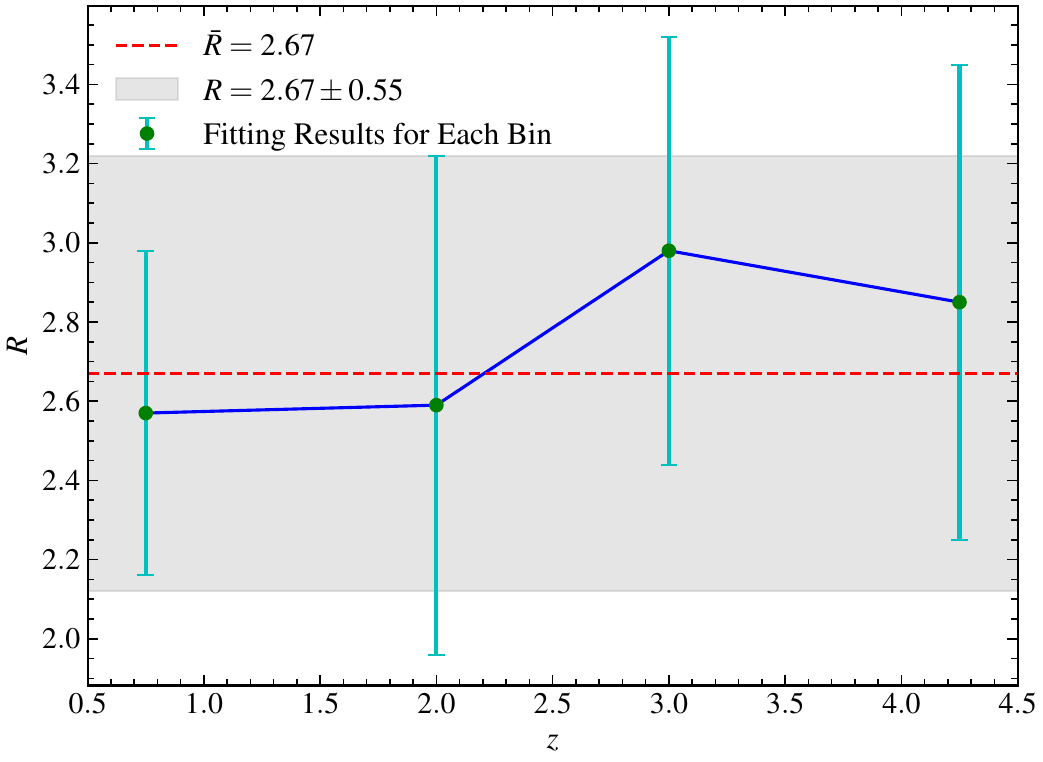}
\caption{The radio-loudness fitting results for the four redshift bins. Green data points represent the means obtained from Gaussian fitting, with error bars indicating the $\sigma_0$ values. The red dashed line and shaded area denote the best fit, as depicted in Figure \ref{fig:Radioloudness}.
\label{fig:Rbar}}
\end{figure}

{The radio-loudness distribution of AGNs shows a bimodal distribution in observations \citep[e.g.][]{1980A&A....88L..12S,1989AJ.....98.1195K,2002AJ....124.2364I,2007ApJ...654...99W}, suggesting the presence of two populations with different radio properties. The distribution can be well fitted using two Gaussian components \citep[e.g.][]{2012ApJ...759...30B,2022PASJ...74..239X}. In this study, we focus on the radio-loud population.} We used the radio-loudness data of 225 radio-loud quasars with redshift from 0 to 5 from \cite{2022MNRAS.513.4673B} to perform a Gaussian fit of the radio-loudness distribution. The definition of radio-loudness is the logarithm of the ratio of the rest-frame luminosity between 5 GHz and $4400\ \AA$ \ ($R={\rm log_{10}}\left(L_{\rm 5GHz} / L_{4400}\right)$). We estimate the radio-loudness distribution of the radio-loud quasars to be
\begin{equation}
N(R)=\frac{1}{\sqrt{2\pi}\sigma_0}\exp\left(\frac{-\left(R-\bar{R}\right)^2}{2\sigma_0^2}\right), \label{eq:NR}
\end{equation} 
where $R$ is radio-loudness, and $\bar{R}$ and $\sigma_0$ are free parameters. Figure \ref{fig:Radioloudness} shows the result of our radio-loudness distribution fitting, with the best-fit result being $\bar{R}=2.67$ and $\sigma_0=0.55$. {We also divided the data into four redshift bins for individual fitting\footnote{The four redshift bins are $[0,1.5]$, $(1.5,2.5]$, $(2.5,3.5]$, and $[3.5,5]$.}, with the results presented in Figure~\ref{fig:Rbar}. As indicated, no significant redshift evolution is observed across the four bins. Separate Kolmogorov-Smirnov test (K-S test) was performed for every two redshift bins. All groups, except for the redshift range $(2.5,3.5]$, passed the K-S test. These results suggest that radio-loudness $R$ either remains constant or exhibits only weak evolution with redshift $z$.} 
Therefore, we believe that applying this distribution form to higher redshifts should be safe.

\subsection{Radio Spectral Index}
To calculate the radio luminosity of a particular band, we assume that the radio spectrum conforms to the following relation,
\begin{equation}
F_{\rm radio} \propto \nu^{\alpha},
\end{equation}
where $F_{\rm radio}$ is the radio flux and $\alpha$ is the radio spectral index. The radio spectral index affects two calculation processes: one involves converting from the observed frame to the rest frame, known as K-correction, and the other is the transformation between the 5 GHz definition of radio-loudness and the SKA-LOW observing band. Based on the findings of \cite{2021A&A...656A.137G}, who derived a median spectral index of $-0.29$ through the stacking of 93 quasars using the Low Frequency Array Two Metre Sky survey (LoTSS-DR2) and the Faint Images of the Radio Sky at Twenty Centimeters (FIRST) data, we adopt $\alpha = -0.29$. {We also note that previous studies of high-redshift radio quasars used a steeper spectral index with $\alpha = -0.75$ \citep{2007AJ....134..617W,2015ApJ...804..118B,2021ApJ...908..124L}, which is in agreement with very long baseline interferometry observations \citep{2005A&A...436L..13F,2008AJ....136..344M,2011A&A...531L...5F,2018ApJ...861...86M}.} Due to our limited understanding of high-redshift quasar radio spectra, we also calculated the impact of $\alpha = -0.75$ on the results as a possible scenario (see Section \ref{subsec:Indeximpact}). 

\begin{figure}
\includegraphics[width=0.47\textwidth]{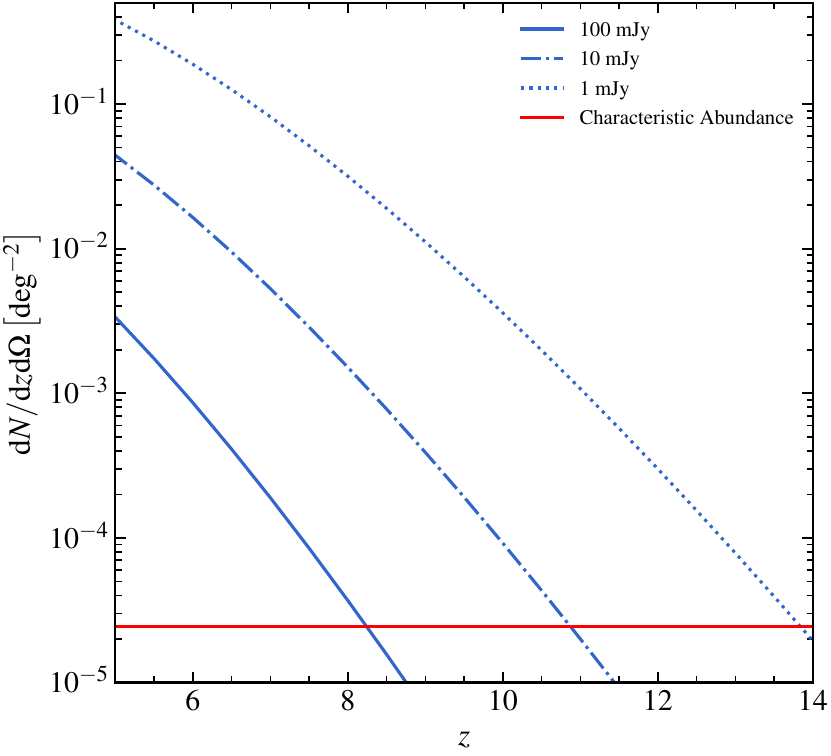}
\caption{Predicted abundance of radio-loud quasars as a function of redshift, with  different flux density thresholds at 150 MHz as indicated in the legend. The red line shows the characteristic abundance ${\Phi}^*$, which corresponds to an abundance where the expected number of observations over the entire sky ($\sim41253\ {\rm deg}^2$) is 1 (${\Phi}^*\simeq 1/41253\ {\rm deg}^{-2}$). Sources with an abundance below this value are almost non-existent.
\label{fig:QAstandard}}
\end{figure}

\subsection{Radio-loud Quasar Abundance}
Combining the above prescriptions, we can get the radio-loud quasar abundance with a radio flux density greater than a certain flux density threshold $F_{\rm threshold}$ as
\begin{alignat}{2}
    &\frac{{\rm d}N}{{\rm d}z{\rm d}\Omega}(F_{\rm threshold}, z)=\frac{{\rm d}V}{{\rm d}z{\rm d}\Omega} D_{\rm q} \nonumber \\
    & \times \quad \int_0^\infty {\rm d}M_{\rm halo} \left(\int_{R_0(M_{\rm halo},F_{\rm threshold})}^\infty {\rm d}R \ N(R) \right) \frac{{\rm d}n}{{\rm d}M_{\rm halo}}, \label{eq:dndm}
\end{alignat}
where ${\rm d}V/{\rm d}z {\rm d}\Omega$ is the cosmological volume element, $R_0$ is the threshold of radio-loudness calculated from the given radio flux density threshold $F_{\rm threshold}$ and halo mass $M_{\rm halo}$, {duty cycle $D_{\rm q}$ can be found in Equation (\ref{eq:Dq}), radio-loudness distribution $N(R)$ can be found in Equation (\ref{eq:NR})}, and ${\rm d}n/{\rm d}M_{\rm halo}$ is the halo abundance calculated from Equations~(\ref{eq:0}) and (\ref{eq:1}).

\section{Results and Discussions}\label{sec:Result and Discuss}

\subsection{Quasar Abundance Predicted by The Model}
Figure \ref{fig:QAstandard} shows the distribution of quasar abundance above a given radio flux density threshold as a function of redshift, obtained according to Equation  (\ref{eq:dndm}). As discussed in Section \ref{sec:Method}, the standard parameter settings we used are as follow: average Eddington ratio $\lambda_{\rm Edd}=1$, average quasar lifetime $t_{\rm q}=10^{5.5}$ years, radio spectral index $\alpha=-0.29$, and radio-loud fraction RLF = 10\%. 

\begin{figure}
\includegraphics[width=0.47\textwidth]{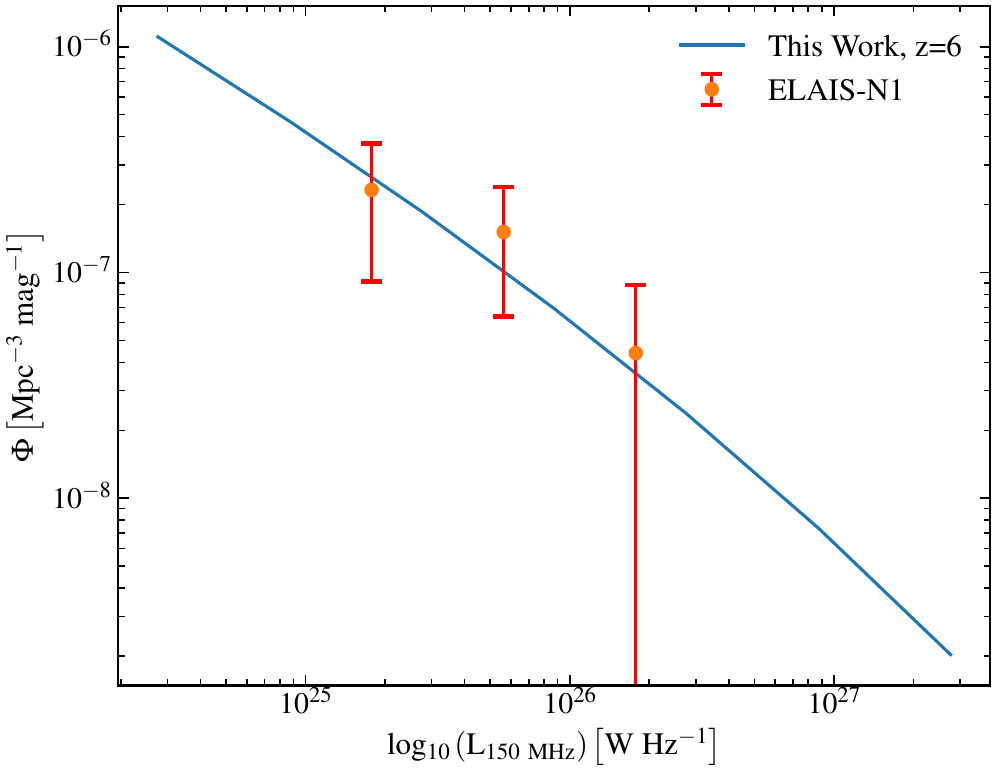}
\caption{Comparison of the model-predicted radio luminosity function with observational data. The blue line represents the radio luminosity function of radio-loud quasars at $z = 6$ calculated by our model, while the red data points are the observational results from the ELAIS-N1 field in the LoTSS Deep Fields DR1, with error bars representing the $1\sigma$ error.
\label{fig:RadioLF}}
\end{figure}

\subsection{Comparison with LOFAR Observations}
We attempt to compare the calibrated model with existing high-redshift radio observations. Specifically, we use data from the {ELAIS-N1} field in the {LoTSS} Deep Fields DR1 \citep{2021A&A...648A...2S,2023MNRAS.523.1729B}, which has the lowest average noise level in the three deep fields. We select radio-excess AGNs with redshifts between 5.6 and 6.4 from this deep field observation data and construct the radio luminosity function using the standard $1/V_{\rm max}$ technique \citep{1968ApJ...151..393S,1989ApJ...338...13C}. The details of the calculation and the incompleteness information can be found in Section 3.1 of \cite{2022MNRAS.513.3742K}. For our model, we convolve the optical luminosity function with the radio-loudness distribution to calculate the radio luminosity function. Figure \ref{fig:RadioLF} shows the comparison of the radio luminosity function predicted by our model with the LOFAR observations. At $z \sim 6$, the model-predicted radio luminosity function matches well with the LOFAR observations. However, we also notice that the observed radio luminosity function is slightly higher than the model prediction (within 1$\sigma$ error range), and this discrepancy is more pronounced in the Lockman Hole field {(about $1.58$$\sigma$)}. Since our model is calibrated using the observed optical luminosity function, this discrepancy may indicate that some high accretion rate AGNs during the reionization epoch are optically obscured, {similar to recently discovered high-redshift obscured AGNs \citep[e.g.][]{2023MNRAS.520.4609E,2024ApJ...961L..25L,2024arXiv240707236B}. This suggests that the optical signals from certain AGNs cannot be effectively observed due to being obscured by dust, while their radio signals can still be detected. Therefore, the results provided by our model are conservative.}

\begin{figure}
\includegraphics[width=0.47\textwidth]{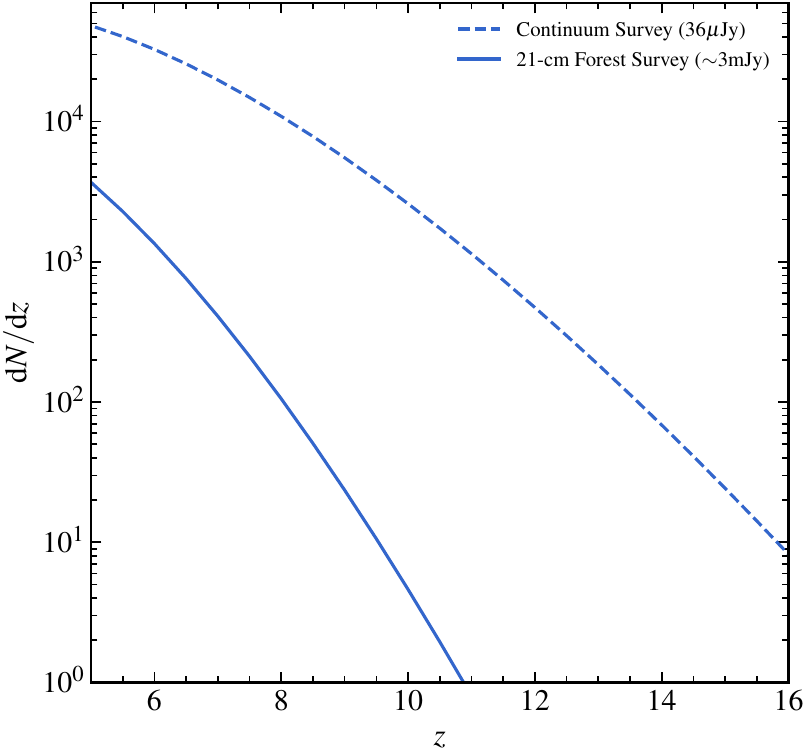}
\caption{The number density of radio-loud quasars as expected in SKA-LOW observations. The dashed line represents the number density for a continuum survey, and the solid line represents the available bright sources for 21-cm forest observations. The sensitivity of the 21-cm forest survey shown in the figure corresponds to the flux density limit at $z=9$, while at $z=6$, the sensitivity is $\sim 1.3$ mJy. The parameters for the continuum survey and the 21-cm forest survey are shown in Table \ref{tab:table1} and Table \ref{tab:table2}, respectively.
\label{fig:SKAstandard}}
\end{figure}

\begin{figure}
\includegraphics[width=0.47\textwidth]{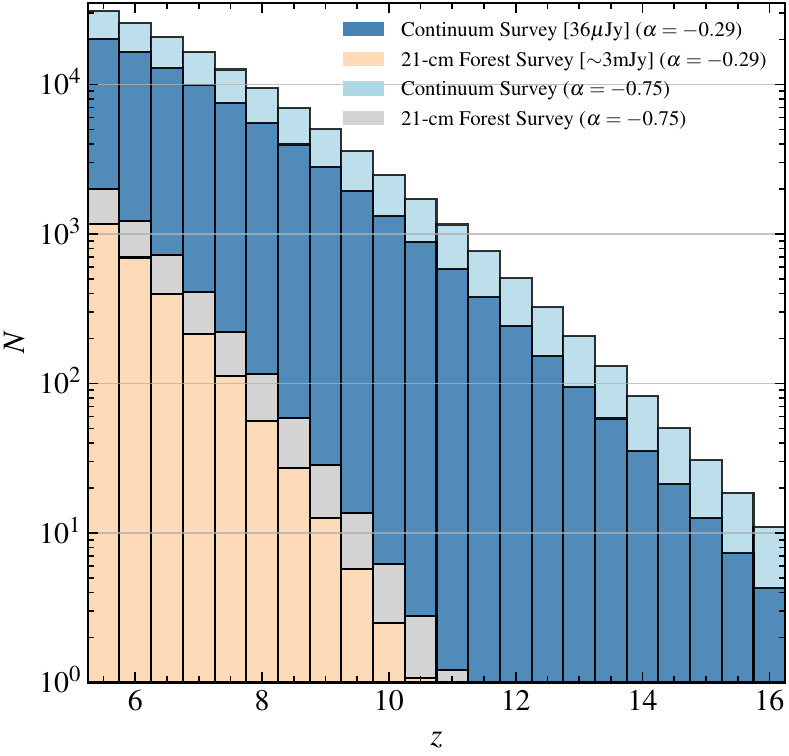}
\caption{The integrated number of radio-loud quasars at various redshifts as expected in SKA-LOW observations. The darker blue and orange histograms are the results of the fiducial model with $\alpha=-0.29$, and the lighter blue and gray histograms represent the case where the radio spectral index $\alpha=-0.75$. The redshift bin width is $0.5$. The sensitivity of the 21-cm forest survey shown in the figure corresponds to the flux density limit at $z=9$, while at $z=6$, the sensitivity is $\sim 1.3$ mJy. 
\label{fig:SKAhistogram}}
\end{figure}

\
\subsection{SKA-LOW Observation Forecast}\label{subsec:SKA}
For SKA-LOW radio-loud quasar observations, we considered two observation modes: the first is a survey to search for quasars. Since the goal of this survey is to find visible radio-loud quasars in the sky, a high frequency resolution is not required, so we refer to it as a continuum survey. The relevant parameters for the continuum survey are summarized in Table \ref{tab:table1}. The second mode is a survey to find radio-loud quasars suitable for 21-cm forest studies. To resolve individual 21-cm lines, this survey requires higher frequency resolution, and thus we call it the 21-cm forest survey. The relevant parameters for the 21-cm forest survey are summarized in Table \ref{tab:table2}. Note that our 21-cm forest survey forecast shows the total number of sources within a survey area of 10313 ${\rm deg}^{2}$ that can resolve individual 21-cm lines with 100~h of integration time. Conducting a 21-cm forest blind survey with this setup would require a substantial amount of observation time ($\sim$ 9 years). From an observational perspective, we should first conduct the continuum blind survey to determine the positions of visible sources and then perform a high frequency-resolution follow-up observations on those bright sources. How to filter high-redshift candidates from the blind survey is an important issue for future observations, {we have performed the relevant discussions in Section \ref{subsec:InfraredLimit}.}

\begin{table}
    \caption{Proposed observation parameters for SKA-LOW Continuum Survey.}
    \label{tab:table1}
    \centering
    \begin{tabular}{p{4cm}ccc}
        \hline\hline
        {Parameter} & {Symbol} & {Value} & {Unit} \\
        \hline
        {Search Coverage} & $S_{\rm tot}$ & 10313 & ${\rm deg}^2$ \\
        {Survey Time} & $T_{\rm tot}$ & $365\times6$ & hour \\
        {Effective Collecting Area} & $A_{\rm{eff}}$ & 419000 & m$^2$ \\
        {Channel Width} & $\delta \nu$ & 10 & MHz \\
        {Antenna Diameter} & $D_{\rm ant}$ & 40 & m \\
        {System Temperature} & $T_{\rm sys}$ & $40+1.1T_{\rm gal}$ & K \\
        \hline
    \end{tabular}
\begin{threeparttable}
\begin{tablenotes}[flushleft]
    \item \textbf{Note.} Instrument parameters are obtained from \cite{2020PASA...37....7S}. The observation time is assumed to be one year, and the effective observation time is 6 hours per day. $T_{\rm gal}=25\left(\frac{408\rm{MHz}}{f(z)}\right)^{2.75}$, where $f(z)$ is the frequency at redshift $z$.
    \end{tablenotes}
\end{threeparttable}
\end{table}

\begin{table}
    \caption{Proposed observation parameters for SKA-LOW 21-cm forest survey.}
    \label{tab:table2}
    \centering
    \begin{tabular}{p{4cm}ccc}
        \hline\hline
        {Parameter} & {Symbol} & {Value} & {Unit} \\
        \hline
        {Sky Coverage} & $S_{\rm tot}$ & 10313 & ${\rm deg}^2$ \\ 
        {Integration Time} & $\delta t$ & 100 & hour \\
        {Effective Collecting Area} & $A_{\rm{eff}}$ & 419000 & m$^2$ \\
        {Channel Width} & $\delta \nu$ & 5 & kHz \\
        {Antenna Diameter} & $D_{\rm ant}$ & 40 & m \\
        {System Temperature} & $T_{\rm sys}$ & $40+1.1T_{\rm gal}$ & K \\
        \hline
    \end{tabular}
\begin{threeparttable}
\begin{tablenotes}[flushleft]
    \item \textbf{Note.} Similar to Table \ref{tab:table1}, the integration time and the channel width vary. See Section \ref{subsec:SKA} for the observation strategy.
    \end{tablenotes}
\end{threeparttable}
\end{table}

To predict the observations of SKA-LOW, we estimate the noise using the following formula \citep{2017isra.book.....T},
\begin{equation}
\delta {S}^{\rm{N}}\approx \frac{2{k}_{\rm{B}}{T}_{\rm{sys}}}{{A}_{\rm{eff}}\sqrt{2\delta \nu \delta t}},
\end{equation}
where $A_{\rm{eff}}$ is the effective collecting area of the telescope, $T_{\rm sys}$ is the system temperature, $\delta \nu$ is the channel width and $\delta t$ is the integration time. For continuum survey, the single-field integration time $\delta t$ is determined by the survey coverage $S_{\rm tot}$ and the total survey time $T_{\rm tot}$,
\begin{equation}
\delta t = \frac{T_{\rm tot}}{S_{\rm tot}/{\mathit\Omega}},
\end{equation}
where ${\mathit\Omega}$ is the field of view, given by ${\mathit\Omega} \simeq \left(1.2\lambda/D_{\rm ant}\right)^2$, in which $D_{\rm ant}$ is the diameter of a station. We estimate ${\mathit\Omega}$ using the wavelength corresponding to 150 MHz.
We require the signal-to-noise ratio to be higher than 5 to detect quasars in a continuum survey, or to detect an absorption line for a follow-up 21-cm observation. For continuum surveys, this means ${\rm flux\ density}>5\times\delta {S}^{\rm{N}}$. For 21-cm forest surveys, we assume an optical depth of $\tau=0.1$, as typical for absorption lines from minihalos or from the IGM with moderate heating during reionization (e.g. \citealt{2006MNRAS.370.1867F,2011MNRAS.410.2025X,2013MNRAS.428.1755C}, which means ${\rm flux\ density} > 5\times\delta {S}^{\rm{N}} /\left(1-e^{-\tau}\right)$.

Figure \ref{fig:SKAstandard} shows the model's predictions for one year of SKA-LOW observations. At redshifts above 9, we expect to observe a large number of radio-loud quasars, among which approximately 20 quasars are bright enough to resolve individual 21-cm lines. This number of radio-loud quasar is enough to make the 21-cm forest an effective probe for the EoR. Note that we assume an effective daily observation time of 6 hours for the survey, which is a conservative estimate. Therefore, the actual number of quasars that can serve as 21-cm forest background sources is expected to be higher than our predictions.

\begin{figure}
\includegraphics[width=0.47\textwidth]{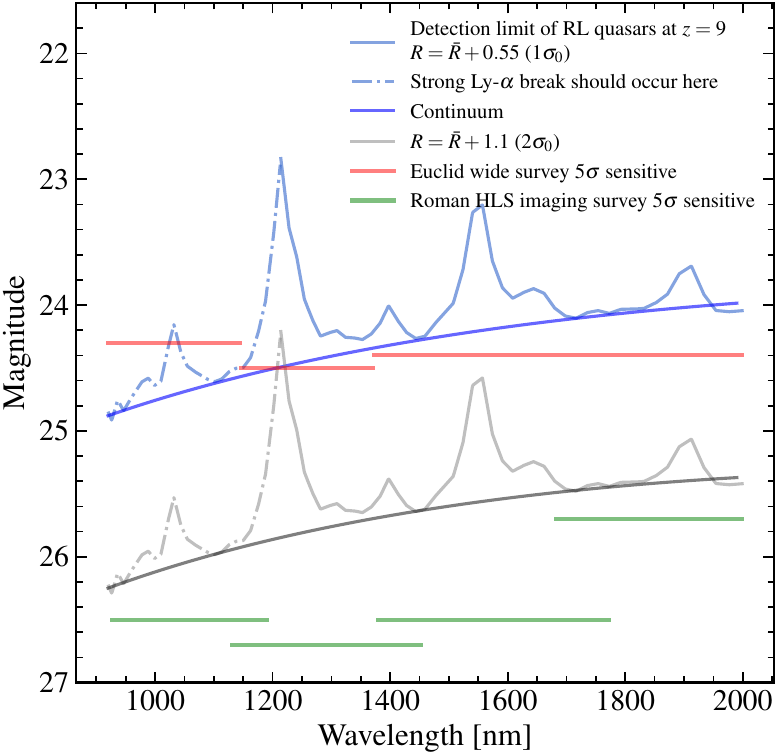}
\caption{The infrared magnitude limit of quasars as required to have a radio flux density exceeding the threshold ($\sim 3\rm mJy$ at $z=9$) for the 21-cm forest survey. The blue and gray lines represent quasars that exceed the average radio-loudness $\bar R$ by $1\sigma_0$ and $2\sigma_0$, respectively. The red and green lines correspond to the $5\sigma$ sensitivity of the Euclid and Roman Space Telescope surveys. The emission line structure is generated by the SED we used. Even quasars with extreme radio-loudness can be detected by future infrared surveys. 
\label{fig:Euclid_Roman}}
\end{figure}

\subsection{Infrared Limit Estimation}\label{subsec:InfraredLimit}

{For the actual observation of high-redshift quasars, obtaining redshift information is crucial. Currently, the search for quasars primarily occurs in the optical and near-infrared bands. Candidate objects are selected through color-color diagram cuts \citep[e.g.][]{2016ApJ...833..222J,2016ApJS..227...11B,2023ApJS..269...27Y,2024ApJ...974..147L}, and redshift values are determined via photometric observations or spectral line identification. For the approximately 20 radio-loud quasars with $z \sim 9$ caculated in Section \ref{subsec:SKA}, the ideal detection method involves infrared-band photometric surveys, which provide redshift data through the identification of the Lyman-$\rm \alpha$ break. While the JWST offers high sensitivity, it is not capable of conducting wide-field survey. In contrast, current and future space-based infrared telescopes, such as Euclid \citep{2022A&A...662A.112E} and the Nancy Grace Roman Space Telescope \citep[Roman Space Telescope,][]{2015arXiv150303757S}, hold significant potential for such surveys.

Based on the radio flux density threshold for the 21-cm forest survey at $z=9$ obtained in Section \ref{subsec:SKA}, we calculated the corresponding infrared AB magnitude threshold by considering radio-loudness and quasar SED. The results, along with the survey sensitivities of Euclid and Roman Space Telescope, are presented in Figure \ref{fig:Euclid_Roman}. Note that the emission line structure is based on the quasar SED we used, and we also fitted the shape of the continuum. As Figure \ref{fig:Euclid_Roman} shows, Euclid can detect quasars with radio-loudness close to $\bar{R} + 1\sigma_0$ over a large survey area ($\sim 14000\ \rm {deg}^2$). In contrast, although the Roman Space Telescope covers a smaller area ($\sim 2200\ \rm {deg}^2$), it can detect a small number of quasars with extreme radio-loudness ($R \gtrsim \bar{R} + 2\sigma_0$). Therefore, we believe that in the SKA era, combined with these infrared space telescopes, we can effectively search for high-redshift radio-loud quasars.}

\begin{figure}
\includegraphics[width=0.47\textwidth]{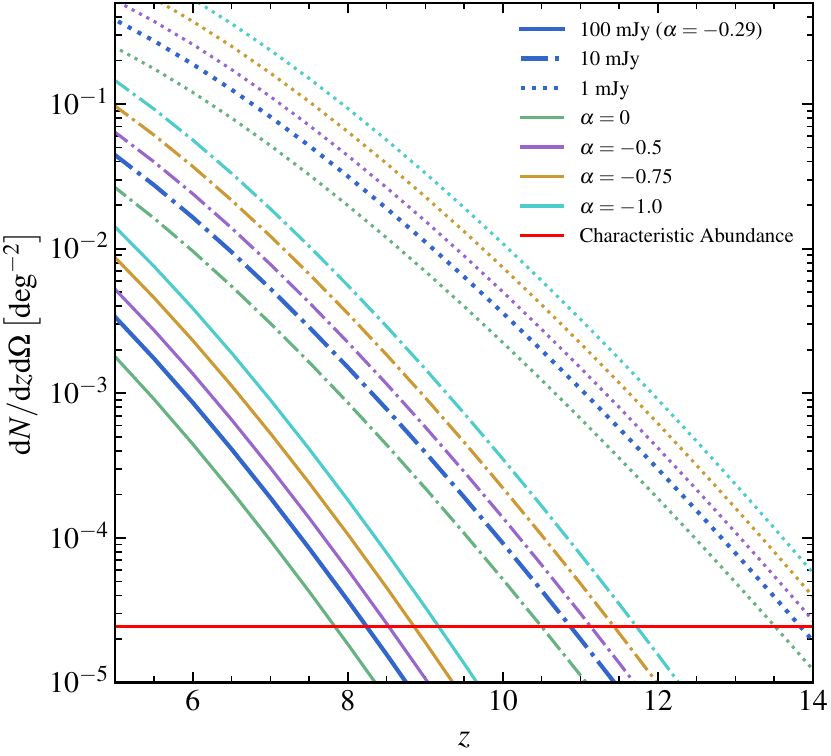}
\caption{Influence of different radio spectral index settings on the abundance of radio-loud quasars. Different colors of lines represent different spectral indices, while different line styles represent different radio flux density thresholds (at 150 MHz). Sources with an abundance below the characteristic abundance marked by the red line are almost non-existent.
\label{fig:QAindex}}
\end{figure}

\begin{figure}
\includegraphics[width=0.47\textwidth]{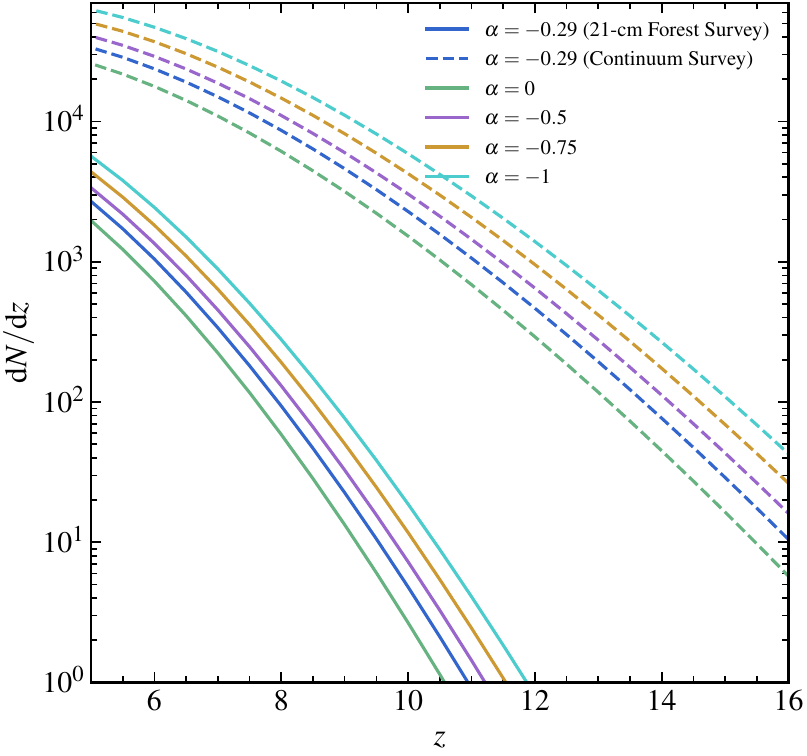}
\caption{Influence of different radio spectral index settings on the forecast results of SKA-LOW observations. The observation frequency band of SKA-LOW is used for calculations here. Specifically, for the continuum surveys, we used 150 MHz, while for the 21-cm forest survey, we used the frequency of the redshifted 21-cm line.
\label{fig:SKAindex}}
\end{figure}

\subsection{Influence of Radio Spectral Index}\label{subsec:Indeximpact}
Figures \ref{fig:QAindex} and \ref{fig:SKAindex} show the influence of different radio spectral indices on the results. Typically, we use $-0.5$ as a standard to determine whether a radio spectrum is steep or flat. Early radio studies of galaxies suggested a steep radio spectral index of $-0.75$ \citep{1992ARA&A..30..575C}, which has been adopted by multiple studies \citep[e.g.,][]{2007AJ....134..617W,2014AJ....147....6M,2015ApJ...804..118B}. {We used the stacking result of \cite{2021A&A...656A.137G}, $\alpha = -0.29$, as the standard parameter setting. However, it is noteworthy that the worse sensitivity of FIRST compared to LoTSS may naturally flatten the obtained spectral index.}
Since our understanding of the radio spectra of quasars is still limited, here we calculate the influence of multiple values of radio spectral indices on the results. {As shown in Figures \ref{fig:QAindex} and \ref{fig:SKAindex}, our model indicates that a steeper spectral index is more beneficial for SKA-LOW's observations of high-redshift radio-loud quasars. This advantage arises because the radio-loudness $R$ we are fitting is defined at $5$ GHz, meaning that our results directly reflect the flux density at this frequency. A steeper radio spectrum produces stronger flux for sources within SKA-LOW's observation frequency range ($50-350$ MHz). The integral results suggest that adopting a steeper spectral index ($\alpha = -0.75$) instead of the flat spectral index proposed by \cite{2021A&A...656A.137G} ($\alpha = -0.29$) could increase the number of observable quasars at $z \sim 9$ by approximately $1.3$ times (from $12$ to $28$). According to the requirements outlined by \cite{2023NatAs...7.1116S} for background sources in the 21-cm forest (10 quasars at $z=9$), we anticipate sufficient background sources can be observed even with a flat spectral index. Consequently, the spectral index is not a critical parameter influencing the success of the 21-cm forest method.}

\subsection{Influence of Radio-Loud Fraction}\label{subsec:RLFimpact}
\begin{figure}
\includegraphics[width=0.47\textwidth]{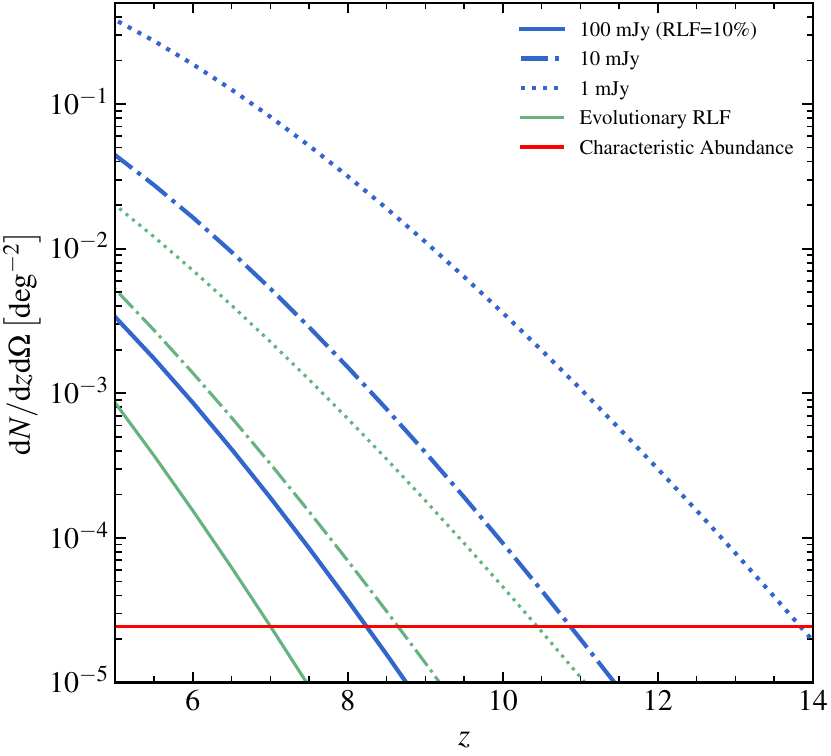}
\caption{Influence of evolving RLF on the abundance of radio-loud quasars. The blue line represents a constant 10\% RLF, while the green line represents an evolving RLF. Different line styles indicate different radio flux density thresholds (at 150 MHz). Sources with an abundance below the characteristic abundance marked by the red line are almost non-existent.
\label{fig:QARLF}}
\end{figure}

Figures \ref{fig:QARLF} and \ref{fig:SKARLF} illustrate the impact of the evolution of RLF with redshift and luminosity on the results. We adopted the evolutionary form from \cite{2007ApJ...656..680J}, which implies that the RLF at high redshift is much lower than 10\%. Under the assumption of evolutionary RLF, the model predicts a significant decrease in the abundance of quasars observable by SKA-LOW's one-year survey, making it more challenging to use the 21-cm forest method to probe cosmic reionization. If this RLF evolution is real, during the reionization period, the number of quasars with 100-hour integration times sufficient to resolve individual 21-cm lines is insufficient to support the 21-cm forest method, and we may need longer observation times to utilize relatively fainter quasars.

\section{conclusion}\label{sec:Conclusion}
One of the interesting results obtained from our model is the average lifetime $t_{\rm q}$ of high-redshift ($z\gtrsim6$) quasars. Assuming Eddington accretion, the observed high-redshift quasar luminosity function requires that the average lifetime of quasars given by the model is $\sim 10^{5.5}$ years (see Section \ref{subsec:Duty Cycle}). Although this result is much smaller than early theoretical estimates ($\sim10^7$ years), it is close to recent constraints using Lyman-$\rm \alpha$ near-zone data. Note that determining the average lifetime of quasars in the model requires a given halo--SMBH mass relationship, and currently there is large uncertainty in the estimation of high-redshift SMBH and halo masses based on quasar spectra. It is foreseeable that with the accumulation of high-redshift observational data, our model can provide a more reliable average lifetime of quasars, which will help us understand the growth process of SMBHs in the early universe.

\begin{figure}
\includegraphics[width=0.47\textwidth]{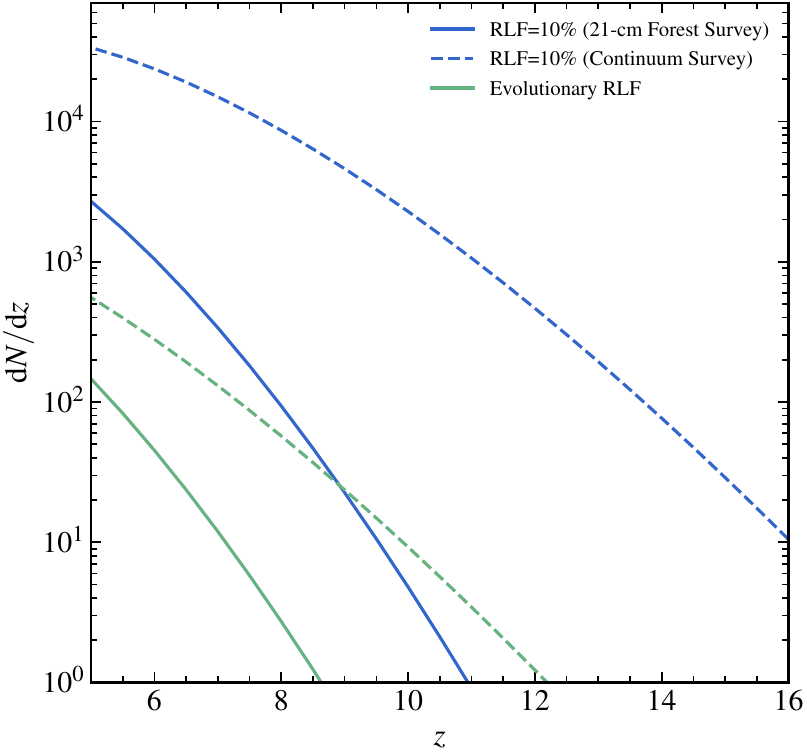}
\caption{Influence of evolving RLF on the forecast results of SKA-LOW observations. The blue line represents a constant 10\% RLF, while the green line represents an evolving RLF. The dash line represents the continuum survey, and the solid line represents the 21-cm forest survey. Detailed descriptions of the continuum survey and the 21-cm forest survey can be found in Section \ref{subsec:SKA}.
\label{fig:SKARLF}}
\end{figure}

Another notable finding concerns the accretion behavior of AGNs. Our model, calibrated at high redshift, {resulted in a short quasar lifetime under the assumption of Eddington accretion}. However, applying this setting to lower redshift results in significant discrepancies with observations. {Additionally, we found that simply changing the quasar lifetime does not help, indicating that the assumption of quasars accreting at the Eddington rate at low redshifts is wrong.} As a simple attempt, we recalibrated the model using the quasar luminosity function at $z = 4$ {and allowed both $\lambda_{\rm Edd}$ and $t_{\rm q}$ to vary simultaneously}, resulting in an average Eddington ratio of $\lambda_{\rm Edd}= 0.04$ and an average quasar lifetime of $t_{\rm q}=10^{8.2}$ years. This suggests that the accretion behaviors of early and late AGNs differ greatly.

The most significant result derived from our model is the predicted abundance of radio-loud quasars observable by the SKA-LOW in an one-year survey. According to the standard parameter settings (Figure \ref{fig:SKAstandard}), we expect to identify over a thousand radio-loud quasars at $z>6$ that are sufficient to resolve individual 21-cm lines, within a year of observation. Even at higher redshifts, such as $z>9$, we can still detect approximately 20 such radio-loud quasars.

We tested the impact of various quasar radio spectral indices on our results (Figure \ref{fig:SKAindex}). Our findings indicate that a steeper radio spectrum is more favorable for observations in the SKA-LOW frequency band. However, the radio spectral index {is not a critical parameter influencing the success of the the 21-cm forest method}.

Our results indicate that whether RLF evolves is crucial for the observation of the 21-cm forest in the SKA era (Figure~\ref{fig:SKARLF}). If the RLF remains close to 10\% at high redshift, then the 21-cm forest can serve as an effective probe of reionization. However, if the RLF evolution claimed by \cite{2007ApJ...656..680J} is real, then it will be challenging to find background sources for detecting individual 21-cm absorption lines, and we may need longer integration times to utilize relatively fainter quasars. In this case, a statistical measurement of the 21-cm forest would be vital in alleviating the requirement on the brightness of background sources \citep{Mack2012,Thyagarajan2020,2023NatAs...7.1116S}.

\section*{acknowledgments}
We thank K. Shimasaku for providing us with the compiled data of the halo--SMBH mass relation. This work was supported by the National SKA Program of China (Grants Nos. 2020SKA0110401, 2022SKA0110200, 2022SKA0110203,  and 2020SKA0110100), the National Natural Science Foundation of China (Grants Nos. 12473001, 11975072, 11875102, and 11835009), the National 111 Project (Grant No. B16009), the CAS Project for Young Scientists in Basic Research (Grant No. YSBR-092), the National Key R\&D Program of China (Grant No. 2022YFF0504300), and the science research grants from the China Manned Space Project (Grants Nos. CMS-CSST-2021-B01 and CMS-CSST-2021-A02). LOFAR data products were provided by the LOFAR Surveys Key Science project (LSKSP; \url{https://lofar-surveys.org/}) and were derived from observations with the International LOFAR Telescope (ILT). LOFAR \citep{2013A&A...556A...2V} is the Low Frequency Array designed and constructed by ASTRON. It has observing, data processing, and data storage facilities in several countries, which are owned by various parties (each with their own funding sources), and which are collectively operated by the ILT foundation under a joint scientific policy. The efforts of the LSKSP have benefited from funding from the European Research Council, NOVA, NWO, CNRS-INSU, the SURF Co-operative, the UK Science and Technology Funding Council and the Jülich Supercomputing Centre.

\bibliography{ref}

\begin{thebibliography}{}
\expandafter\ifx\csname natexlab\endcsname\relax\def\natexlab#1{#1}\fi
\providecommand{\url}[1]{\href{#1}{#1}}
\providecommand{\dodoi}[1]{doi:~\href{http://doi.org/#1}{\nolinkurl{#1}}}
\providecommand{\doeprint}[1]{\href{http://ascl.net/#1}{\nolinkurl{http://ascl.net/#1}}}
\providecommand{\doarXiv}[1]{\href{https://arxiv.org/abs/#1}{\nolinkurl{https://arxiv.org/abs/#1}}}

\bibitem[{{Ba{\~n}ados} {et~al.}(2015){Ba{\~n}ados}, {Venemans}, {Morganson},
  {Hodge}, {Decarli}, {Walter}, {Stern}, {Schlafly}, {Farina}, {Greiner},
  {Chambers}, {Fan}, {Rix}, {Burgett}, {Draper}, {Flewelling}, {Kaiser},
  {Metcalfe}, {Morgan}, {Tonry}, \& {Wainscoat}}]{2015ApJ...804..118B}
{Ba{\~n}ados}, E., {Venemans}, B.~P., {Morganson}, E., {et~al.} 2015, \apj,
  804, 118, \dodoi{10.1088/0004-637X/804/2/118}

\bibitem[{{Ba{\~n}ados} {et~al.}(2016){Ba{\~n}ados}, {Venemans}, {Decarli},
  {Farina}, {Mazzucchelli}, {Walter}, {Fan}, {Stern}, {Schlafly}, {Chambers},
  {Rix}, {Jiang}, {McGreer}, {Simcoe}, {Wang}, {Yang}, {Morganson}, {De Rosa},
  {Greiner}, {Balokovi{\'c}}, {Burgett}, {Cooper}, {Draper}, {Flewelling},
  {Hodapp}, {Jun}, {Kaiser}, {Kudritzki}, {Magnier}, {Metcalfe}, {Miller},
  {Schindler}, {Tonry}, {Wainscoat}, {Waters}, \& {Yang}}]{2016ApJS..227...11B}
{Ba{\~n}ados}, E., {Venemans}, B.~P., {Decarli}, R., {et~al.} 2016, \apjs, 227,
  11, \dodoi{10.3847/0067-0049/227/1/11}

\bibitem[{{Ba{\~n}ados} {et~al.}(2021){Ba{\~n}ados}, {Mazzucchelli}, {Momjian},
  {Eilers}, {Wang}, {Schindler}, {Connor}, {Andika}, {Barth}, {Carilli},
  {Davies}, {Decarli}, {Fan}, {Farina}, {Hennawi}, {Pensabene}, {Stern},
  {Venemans}, {Wenzl}, \& {Yang}}]{2021ApJ...909...80B}
{Ba{\~n}ados}, E., {Mazzucchelli}, C., {Momjian}, E., {et~al.} 2021, \apj, 909,
  80, \dodoi{10.3847/1538-4357/abe239}

\bibitem[{{Ba{\~n}ados} {et~al.}(2023){Ba{\~n}ados}, {Schindler}, {Venemans},
  {Connor}, {Decarli}, {Farina}, {Mazzucchelli}, {Meyer}, {Stern}, {Walter},
  {Fan}, {Hennawi}, {Khusanova}, {Morrell}, {Nanni}, {Noirot}, {Pensabene},
  {Rix}, {Simon}, {Verdoes Kleijn}, {Xie}, {Yang}, \&
  {Connor}}]{2023ApJS..265...29B}
{Ba{\~n}ados}, E., {Schindler}, J.-T., {Venemans}, B.~P., {et~al.} 2023, \apjs,
  265, 29, \dodoi{10.3847/1538-4365/acb3c7}

\bibitem[{{Balokovi{\'c}} {et~al.}(2012){Balokovi{\'c}}, {Smol{\v{c}}i{\'c}},
  {Ivezi{\'c}}, {Zamorani}, {Schinnerer}, \& {Kelly}}]{2012ApJ...759...30B}
{Balokovi{\'c}}, M., {Smol{\v{c}}i{\'c}}, V., {Ivezi{\'c}}, {\v{Z}}., {et~al.}
  2012, \apj, 759, 30, \dodoi{10.1088/0004-637X/759/1/30}

\bibitem[{{Banados} {et~al.}(2024){Banados}, {Momjian}, {Connor}, {Belladitta},
  {Decarli}, {Mazzucchelli}, {Venemans}, {Walter}, {Wang}, {Xie}, {Barth},
  {Eilers}, {Fan}, {Khusanova}, {Schindler}, {Stern}, {Yang}, {Taufik Andika},
  {Carilli}, {Farina}, {Fabian}, {Hennawi}, {Pensabene}, \&
  {Rojas-Ruiz}}]{2024arXiv240707236B}
{Banados}, E., {Momjian}, E., {Connor}, T., {et~al.} 2024, arXiv e-prints,
  arXiv:2407.07236, \dodoi{10.48550/arXiv.2407.07236}

\bibitem[{Bariuan {et~al.}(2022)Bariuan, Snios, Sobolewska, Siemiginowska, \&
  Schwartz}]{2022MNRAS.513.4673B}
Bariuan, L. G.~C., Snios, B., Sobolewska, M., Siemiginowska, A., \& Schwartz,
  D.~A. 2022, \mnras, 513, 4673, \dodoi{10.1093/mnras/stac1153}

\bibitem[{{Best} {et~al.}(2023){Best}, {Kondapally}, {Williams}, {Cochrane},
  {Duncan}, {Hale}, {Haskell}, {Ma{\l}ek}, {McCheyne}, {Smith}, {Wang},
  {Botteon}, {Bonato}, {Bondi}, {Calistro Rivera}, {Gao}, {G{\"u}rkan},
  {Hardcastle}, {Jarvis}, {Mingo}, {Miraghaei}, {Morabito}, {Nisbet},
  {Prandoni}, {R{\"o}ttgering}, {Sabater}, {Shimwell}, {Tasse}, \& {van
  Weeren}}]{2023MNRAS.523.1729B}
{Best}, P.~N., {Kondapally}, R., {Williams}, W.~L., {et~al.} 2023, \mnras, 523,
  1729, \dodoi{10.1093/mnras/stad1308}

\bibitem[{{Bolgar} {et~al.}(2018){Bolgar}, {Eames}, {Hottier}, \&
  {Semelin}}]{2018MNRAS.478.5564B}
{Bolgar}, F., {Eames}, E., {Hottier}, C., \& {Semelin}, B. 2018, \mnras, 478,
  5564, \dodoi{10.1093/mnras/sty1293}

\bibitem[{{Bowman} {et~al.}(2018){Bowman}, {Rogers}, {Monsalve}, {Mozdzen}, \&
  {Mahesh}}]{2018Natur.555...67B}
{Bowman}, J.~D., {Rogers}, A. E.~E., {Monsalve}, R.~A., {Mozdzen}, T.~J., \&
  {Mahesh}, N. 2018, \nat, 555, 67, \dodoi{10.1038/nature25792}

\bibitem[{{Carilli} {et~al.}(2002){Carilli}, {Gnedin}, \&
  {Owen}}]{2002ApJ...577...22C}
{Carilli}, C.~L., {Gnedin}, N.~Y., \& {Owen}, F. 2002, \apj, 577, 22,
  \dodoi{10.1086/342179}

\bibitem[{{Chatterjee} {et~al.}(2024){Chatterjee}, {Elahi}, {Bharadwaj},
  {Sarkar}, {Choudhuri}, {Sethi}, \& {Patwa}}]{2024arXiv240510080C}
{Chatterjee}, S., {Elahi}, K. M.~A., {Bharadwaj}, S., {et~al.} 2024, arXiv
  e-prints, arXiv:2405.10080, \dodoi{10.48550/arXiv.2405.10080}

\bibitem[{{Chen} {et~al.}(2023){Chen}, {Yan}, {Xu}, {Deng}, {Wu}, {Wu}, {Zhou},
  {Zhang}, {Zhu}, {Yang}, \& {Wu}}]{2023ChJSS..43...43C}
{Chen}, X., {Yan}, J., {Xu}, Y., {et~al.} 2023, ChJSS, 43, 43,
  \dodoi{10.11728/cjss2023.01.220104001}

\bibitem[{{Ciardi} {et~al.}(2015){Ciardi}, {Inoue}, {Mack}, {Xu}, \&
  {Bernardi}}]{2015aska.confE...6C}
{Ciardi}, B., {Inoue}, S., {Mack}, K., {Xu}, Y., \& {Bernardi}, G. 2015, in
  Advancing Astrophysics with the Square Kilometre Array (AASKA14), 6,
  \dodoi{10.22323/1.215.0006}

\bibitem[{{Ciardi} {et~al.}(2013){Ciardi}, {Labropoulos}, {Maselli}, {Thomas},
  {Zaroubi}, {Graziani}, {Bolton}, {Bernardi}, {Brentjens}, {de Bruyn},
  {Daiboo}, {Harker}, {Jelic}, {Kazemi}, {Koopmans}, {Martinez}, {Mellema},
  {Offringa}, {Pandey}, {Schaye}, {Veligatla}, {Vedantham}, \&
  {Yatawatta}}]{2013MNRAS.428.1755C}
{Ciardi}, B., {Labropoulos}, P., {Maselli}, A., {et~al.} 2013, \mnras, 428,
  1755, \dodoi{10.1093/mnras/sts156}

\bibitem[{{Condon}(1989)}]{1989ApJ...338...13C}
{Condon}, J.~J. 1989, \apj, 338, 13, \dodoi{10.1086/167176}

\bibitem[{{Condon}(1992)}]{1992ARA&A..30..575C}
---. 1992, \araa, 30, 575, \dodoi{10.1146/annurev.aa.30.090192.003043}

\bibitem[{{de Lera Acedo} {et~al.}(2022){de Lera Acedo}, {de Villiers},
  {Razavi-Ghods}, {Handley}, {Fialkov}, {Magro}, {Anstey}, {Bevins}, {Chiello},
  {Cumner}, {Josaitis}, {Roque}, {Sims}, {Scheutwinkel}, {Alexander},
  {Bernardi}, {Carey}, {Cavillot}, {Croukamp}, {Ely}, {Gessey-Jones},
  {Gueuning}, {Hills}, {Kulkarni}, {Maiolino}, {Meerburg}, {Mittal},
  {Pritchard}, {Puchwein}, {Saxena}, {Shen}, {Smirnov}, {Spinelli}, \&
  {Zarb-Adami}}]{2022NatAs...6..984D}
{de Lera Acedo}, E., {de Villiers}, D.~I.~L., {Razavi-Ghods}, N., {et~al.}
  2022, NatAs, 6, 984, \dodoi{10.1038/s41550-022-01709-9}

\bibitem[{{DeBoer} {et~al.}(2017){DeBoer}, {Parsons}, {Aguirre}, {Alexander},
  {Ali}, {Beardsley}, {Bernardi}, {Bowman}, {Bradley}, {Carilli}, {Cheng}, {de
  Lera Acedo}, {Dillon}, {Ewall-Wice}, {Fadana}, {Fagnoni}, {Fritz},
  {Furlanetto}, {Glendenning}, {Greig}, {Grobbelaar}, {Hazelton}, {Hewitt},
  {Hickish}, {Jacobs}, {Julius}, {Kariseb}, {Kohn}, {Lekalake}, {Liu}, {Loots},
  {MacMahon}, {Malan}, {Malgas}, {Maree}, {Martinot}, {Mathison}, {Matsetela},
  {Mesinger}, {Morales}, {Neben}, {Patra}, {Pieterse}, {Pober}, {Razavi-Ghods},
  {Ringuette}, {Robnett}, {Rosie}, {Sell}, {Smith}, {Syce}, {Tegmark},
  {Thyagarajan}, {Williams}, \& {Zheng}}]{2017PASP..129d5001D}
{DeBoer}, D.~R., {Parsons}, A.~R., {Aguirre}, J.~E., {et~al.} 2017, \pasp, 129,
  045001, \dodoi{10.1088/1538-3873/129/974/045001}

\bibitem[{{Eilers} {et~al.}(2017){Eilers}, {Davies}, {Hennawi}, {Prochaska},
  {Luki{\'c}}, \& {Mazzucchelli}}]{2017ApJ...840...24E}
{Eilers}, A.-C., {Davies}, F.~B., {Hennawi}, J.~F., {et~al.} 2017, \apj, 840,
  24, \dodoi{10.3847/1538-4357/aa6c60}

\bibitem[{{Eilers} {et~al.}(2021){Eilers}, {Hennawi}, {Davies}, \&
  {Simcoe}}]{2021ApJ...917...38E}
{Eilers}, A.-C., {Hennawi}, J.~F., {Davies}, F.~B., \& {Simcoe}, R.~A. 2021,
  \apj, 917, 38, \dodoi{10.3847/1538-4357/ac0a76}

\bibitem[{{Eisenstein} \& {Hu}(1999)}]{1999ApJ...511....5E}
{Eisenstein}, D.~J., \& {Hu}, W. 1999, \apj, 511, 5, \dodoi{10.1086/306640}

\bibitem[{{Endsley} {et~al.}(2023){Endsley}, {Stark}, {Lyu}, {Wang}, {Yang},
  {Fan}, {Smit}, {Bouwens}, {Hainline}, \& {Schouws}}]{2023MNRAS.520.4609E}
{Endsley}, R., {Stark}, D.~P., {Lyu}, J., {et~al.} 2023, \mnras, 520, 4609,
  \dodoi{10.1093/mnras/stad266}

\bibitem[{{Euclid Collaboration} {et~al.}(2022){Euclid Collaboration},
  {Scaramella}, {Amiaux}, {Mellier}, {Burigana}, {Carvalho}, {Cuillandre}, {Da
  Silva}, {Derosa}, {Dinis}, {Maiorano}, {Maris}, {Tereno}, {Laureijs},
  {Boenke}, {Buenadicha}, {Dupac}, {Gaspar Venancio}, {G{\'o}mez-{\'A}lvarez},
  {Hoar}, {Lorenzo Alvarez}, {Racca}, {Saavedra-Criado}, {Schwartz}, {Vavrek},
  {Schirmer}, {Aussel}, {Azzollini}, {Cardone}, {Cropper}, {Ealet}, {Garilli},
  {Gillard}, {Granett}, {Guzzo}, {Hoekstra}, {Jahnke}, {Kitching}, {Maciaszek},
  {Meneghetti}, {Miller}, {Nakajima}, {Niemi}, {Pasian}, {Percival},
  {Pottinger}, {Sauvage}, {Scodeggio}, {Wachter}, {Zacchei}, {Aghanim},
  {Amara}, {Auphan}, {Auricchio}, {Awan}, {Balestra}, {Bender}, {Bodendorf},
  {Bonino}, {Branchini}, {Brau-Nogue}, {Brescia}, {Candini}, {Capobianco},
  {Carbone}, {Carlberg}, {Carretero}, {Casas}, {Castander}, {Castellano},
  {Cavuoti}, {Cimatti}, {Cledassou}, {Congedo}, {Conselice}, {Conversi},
  {Copin}, {Corcione}, {Costille}, {Courbin}, {Degaudenzi}, {Douspis},
  {Dubath}, {Duncan}, {Dusini}, {Farrens}, {Ferriol}, {Fosalba}, {Fourmanoit},
  {Frailis}, {Franceschi}, {Franzetti}, {Fumana}, {Gillis}, {Giocoli},
  {Grazian}, {Grupp}, {Haugan}, {Holmes}, {Hormuth}, {Hudelot}, {Kermiche},
  {Kiessling}, {Kilbinger}, {Kohley}, {Kubik}, {K{\"u}mmel}, {Kunz},
  {Kurki-Suonio}, {Lahav}, {Ligori}, {Lilje}, {Lloro}, {Mansutti}, {Marggraf},
  {Markovic}, {Marulli}, {Massey}, {Maurogordato}, {Melchior}, {Merlin},
  {Meylan}, {Mohr}, {Moresco}, {Morin}, {Moscardini}, {Munari}, {Nichol},
  {Padilla}, {Paltani}, {Peacock}, {Pedersen}, {Pettorino}, {Pires}, {Poncet},
  {Popa}, {Pozzetti}, {Raison}, {Rebolo}, {Rhodes}, {Rix}, {Roncarelli},
  {Rossetti}, {Saglia}, {Schneider}, {Schrabback}, {Secroun}, {Seidel},
  {Serrano}, {Sirignano}, {Sirri}, {Skottfelt}, {Stanco}, {Starck},
  {Tallada-Cresp{\'\i}}, {Tavagnacco}, {Taylor}, {Teplitz}, {Toledo-Moreo},
  {Torradeflot}, {Trifoglio}, {Valentijn}, {Valenziano}, {Verdoes Kleijn},
  {Wang}, {Welikala}, {Weller}, {Wetzstein}, {Zamorani}, {Zoubian}, {Andreon},
  {Baldi}, {Bardelli}, {Boucaud}, {Camera}, {Di Ferdinando}, {Fabbian},
  {Farinelli}, {Galeotta}, {Graci{\'a}-Carpio}, {Maino}, {Medinaceli}, {Mei},
  {Neissner}, {Polenta}, {Renzi}, {Romelli}, {Rosset}, {Sureau}, {Tenti},
  {Vassallo}, {Zucca}, {Baccigalupi}, {Balaguera-Antol{\'\i}nez}, {Battaglia},
  {Biviano}, {Borgani}, {Bozzo}, {Cabanac}, {Cappi}, {Casas}, {Castignani},
  {Colodro-Conde}, {Coupon}, {Courtois}, {Cuby}, {de la Torre}, {Desai},
  {Dole}, {Fabricius}, {Farina}, {Ferreira}, {Finelli}, {Flose-Reimberg},
  {Fotopoulou}, {Ganga}, {Gozaliasl}, {Hook}, {Keihanen}, {Kirkpatrick},
  {Liebing}, {Lindholm}, {Mainetti}, {Martinelli}, {Martinet}, {Maturi},
  {McCracken}, {Metcalf}, {Morgante}, {Nightingale}, {Nucita}, {Patrizii},
  {Potter}, {Riccio}, {S{\'a}nchez}, {Sapone}, {Schewtschenko}, {Schultheis},
  {Scottez}, {Teyssier}, {Tutusaus}, {Valiviita}, {Viel}, {Vriend}, \&
  {Whittaker}}]{2022A&A...662A.112E}
{Euclid Collaboration}, {Scaramella}, R., {Amiaux}, J., {et~al.} 2022, \aap,
  662, A112, \dodoi{10.1051/0004-6361/202141938}

\bibitem[{{Ferrarese}(2002)}]{2002ApJ...578...90F}
{Ferrarese}, L. 2002, \apj, 578, 90, \dodoi{10.1086/342308}

\bibitem[{{Ferrarese} \& {Merritt}(2000)}]{2000ApJ...539L...9F}
{Ferrarese}, L., \& {Merritt}, D. 2000, \apjl, 539, L9, \dodoi{10.1086/312838}

\bibitem[{{Frey} {et~al.}(2011){Frey}, {Paragi}, {Gurvits}, {Gab{\'a}nyi}, \&
  {Cseh}}]{2011A&A...531L...5F}
{Frey}, S., {Paragi}, Z., {Gurvits}, L.~I., {Gab{\'a}nyi}, K.~{\'E}., \&
  {Cseh}, D. 2011, \aap, 531, L5, \dodoi{10.1051/0004-6361/201117341}

\bibitem[{{Frey} {et~al.}(2005){Frey}, {Paragi}, {Mosoni}, \&
  {Gurvits}}]{2005A&A...436L..13F}
{Frey}, S., {Paragi}, Z., {Mosoni}, L., \& {Gurvits}, L.~I. 2005, \aap, 436,
  L13, \dodoi{10.1051/0004-6361:200500112}

\bibitem[{{Furlanetto}(2006)}]{2006MNRAS.370.1867F}
{Furlanetto}, S.~R. 2006, \mnras, 370, 1867,
  \dodoi{10.1111/j.1365-2966.2006.10603.x}

\bibitem[{{Furlanetto} \& {Loeb}(2002)}]{2002ApJ...579....1F}
{Furlanetto}, S.~R., \& {Loeb}, A. 2002, \apj, 579, 1, \dodoi{10.1086/342757}

\bibitem[{{Furlanetto} {et~al.}(2006){Furlanetto}, {Oh}, \&
  {Briggs}}]{2006PhR...433..181F}
{Furlanetto}, S.~R., {Oh}, S.~P., \& {Briggs}, F.~H. 2006, \physrep, 433, 181,
  \dodoi{10.1016/j.physrep.2006.08.002}

\bibitem[{{Gebhardt} {et~al.}(2000){Gebhardt}, {Bender}, {Bower}, {Dressler},
  {Faber}, {Filippenko}, {Green}, {Grillmair}, {Ho}, {Kormendy}, {Lauer},
  {Magorrian}, {Pinkney}, {Richstone}, \& {Tremaine}}]{2000ApJ...539L..13G}
{Gebhardt}, K., {Bender}, R., {Bower}, G., {et~al.} 2000, \apjl, 539, L13,
  \dodoi{10.1086/312840}

\bibitem[{{Gloudemans} {et~al.}(2021){Gloudemans}, {Duncan}, {R{\"o}ttgering},
  {Shimwell}, {Venemans}, {Best}, {Br{\"u}ggen}, {Calistro Rivera}, {Drabent},
  {Hardcastle}, {Miley}, {Schwarz}, {Saxena}, {Smith}, \&
  {Williams}}]{2021A&A...656A.137G}
{Gloudemans}, A.~J., {Duncan}, K.~J., {R{\"o}ttgering}, H.~J.~A., {et~al.}
  2021, \aap, 656, A137, \dodoi{10.1051/0004-6361/202141722}

\bibitem[{{Gloudemans} {et~al.}(2022){Gloudemans}, {Duncan}, {Saxena},
  {Harikane}, {Hill}, {Zeimann}, {R{\"o}ttgering}, {Yang}, {Best},
  {Ba{\~n}ados}, {Drabent}, {Hardcastle}, {Hennawi}, {Lansbury},
  {Magliocchetti}, {Miley}, {Nanni}, {Shimwell}, {Smith}, {Venemans}, \&
  {Wagenveld}}]{2022A&A...668A..27G}
{Gloudemans}, A.~J., {Duncan}, K.~J., {Saxena}, A., {et~al.} 2022, \aap, 668,
  A27, \dodoi{10.1051/0004-6361/202244763}

\bibitem[{{Goulding} {et~al.}(2023){Goulding}, {Greene}, {Setton}, {Labbe},
  {Bezanson}, {Miller}, {Atek}, {Bogd{\'a}n}, {Brammer}, {Chemerynska},
  {Cutler}, {Dayal}, {Fudamoto}, {Fujimoto}, {Furtak}, {Kokorev}, {Khullar},
  {Leja}, {Marchesini}, {Natarajan}, {Nelson}, {Oesch}, {Pan}, {Papovich},
  {Price}, {van Dokkum}, {Wang}, {Weaver}, {Whitaker}, \&
  {Zitrin}}]{2023ApJ...955L..24G}
{Goulding}, A.~D., {Greene}, J.~E., {Setton}, D.~J., {et~al.} 2023, \apjl, 955,
  L24, \dodoi{10.3847/2041-8213/acf7c5}

\bibitem[{{G{\"u}ltekin} {et~al.}(2009){G{\"u}ltekin}, {Richstone}, {Gebhardt},
  {Lauer}, {Tremaine}, {Aller}, {Bender}, {Dressler}, {Faber}, {Filippenko},
  {Green}, {Ho}, {Kormendy}, {Magorrian}, {Pinkney}, \&
  {Siopis}}]{2009ApJ...698..198G}
{G{\"u}ltekin}, K., {Richstone}, D.~O., {Gebhardt}, K., {et~al.} 2009, \apj,
  698, 198, \dodoi{10.1088/0004-637X/698/1/198}

\bibitem[{{Gupta} {et~al.}(2017){Gupta}, {Ajithkumar}, {Kale}, {Nayak},
  {Sabhapathy}, {Sureshkumar}, {Swami}, {Chengalur}, {Ghosh},
  {Ishwara-Chandra}, {Joshi}, {Kanekar}, {Lal}, \& {Roy}}]{2017CSci..113..707G}
{Gupta}, Y., {Ajithkumar}, B., {Kale}, H.~S., {et~al.} 2017, CSic, 113, 707,
  \dodoi{10.18520/cs/v113/i04/707-714}

\bibitem[{{Haehnelt} {et~al.}(1998){Haehnelt}, {Natarajan}, \&
  {Rees}}]{1998MNRAS.300..817H}
{Haehnelt}, M.~G., {Natarajan}, P., \& {Rees}, M.~J. 1998, \mnras, 300, 817,
  \dodoi{10.1046/j.1365-8711.1998.01951.x}

\bibitem[{{Haiman} \& {Loeb}(1998)}]{1998ApJ...503..505H}
{Haiman}, Z., \& {Loeb}, A. 1998, \apj, 503, 505, \dodoi{10.1086/306017}

\bibitem[{{Haiman} {et~al.}(2004){Haiman}, {Quataert}, \&
  {Bower}}]{2004ApJ...612..698H}
{Haiman}, Z., {Quataert}, E., \& {Bower}, G.~C. 2004, \apj, 612, 698,
  \dodoi{10.1086/422834}

\bibitem[{{Harikane} {et~al.}(2023){Harikane}, {Zhang}, {Nakajima}, {Ouchi},
  {Isobe}, {Ono}, {Hatano}, {Xu}, \& {Umeda}}]{2023ApJ...959...39H}
{Harikane}, Y., {Zhang}, Y., {Nakajima}, K., {et~al.} 2023, \apj, 959, 39,
  \dodoi{10.3847/1538-4357/ad029e}

\bibitem[{{Huynh} \& {Lazio}(2013)}]{2013arXiv1311.4288H}
{Huynh}, M., \& {Lazio}, J. 2013, arXiv e-prints, arXiv:1311.4288,
  \dodoi{10.48550/arXiv.1311.4288}

\bibitem[{{Ighina} {et~al.}(2021){Ighina}, {Belladitta}, {Caccianiga},
  {Broderick}, {Drouart}, {Moretti}, \& {Seymour}}]{2021A&A...647L..11I}
{Ighina}, L., {Belladitta}, S., {Caccianiga}, A., {et~al.} 2021, \aap, 647,
  L11, \dodoi{10.1051/0004-6361/202140362}

\bibitem[{{Ighina} {et~al.}(2023){Ighina}, {Caccianiga}, {Moretti},
  {Belladitta}, {Broderick}, {Drouart}, {Leung}, \&
  {Seymour}}]{2023MNRAS.519.2060I}
{Ighina}, L., {Caccianiga}, A., {Moretti}, A., {et~al.} 2023, \mnras, 519,
  2060, \dodoi{10.1093/mnras/stac3668}

\bibitem[{{Ivezi{\'c}} {et~al.}(2002){Ivezi{\'c}}, {Menou}, {Knapp}, {Strauss},
  {Lupton}, {Vanden Berk}, {Richards}, {Tremonti}, {Weinstein}, {Anderson},
  {Bahcall}, {Becker}, {Bernardi}, {Blanton}, {Eisenstein}, {Fan},
  {Finkbeiner}, {Finlator}, {Frieman}, {Gunn}, {Hall}, {Kim}, {Kinkhabwala},
  {Narayanan}, {Rockosi}, {Schlegel}, {Schneider}, {Strateva}, {SubbaRao},
  {Thakar}, {Voges}, {White}, {Yanny}, {Brinkmann}, {Doi}, {Fukugita},
  {Hennessy}, {Munn}, {Nichol}, \& {York}}]{2002AJ....124.2364I}
{Ivezi{\'c}}, {\v{Z}}., {Menou}, K., {Knapp}, G.~R., {et~al.} 2002, \aj, 124,
  2364, \dodoi{10.1086/344069}

\bibitem[{{Jiang} {et~al.}(2007){Jiang}, {Fan}, {Ivezi{\'c}}, {Richards},
  {Schneider}, {Strauss}, \& {Kelly}}]{2007ApJ...656..680J}
{Jiang}, L., {Fan}, X., {Ivezi{\'c}}, {\v{Z}}., {et~al.} 2007, \apj, 656, 680,
  \dodoi{10.1086/510831}

\bibitem[{{Jiang} {et~al.}(2016){Jiang}, {McGreer}, {Fan}, {Strauss},
  {Ba{\~n}ados}, {Becker}, {Bian}, {Farnsworth}, {Shen}, {Wang}, {Wang},
  {Wang}, {White}, {Wu}, {Wu}, {Yang}, \& {Yang}}]{2016ApJ...833..222J}
{Jiang}, L., {McGreer}, I.~D., {Fan}, X., {et~al.} 2016, \apj, 833, 222,
  \dodoi{10.3847/1538-4357/833/2/222}

\bibitem[{{Kadota} {et~al.}(2021){Kadota}, {Sekiguchi}, \&
  {Tashiro}}]{2021PhRvD.103b3521K}
{Kadota}, K., {Sekiguchi}, T., \& {Tashiro}, H. 2021, \prd, 103, 023521,
  \dodoi{10.1103/PhysRevD.103.023521}

\bibitem[{{Keller} {et~al.}(2024){Keller}, {Thyagarajan}, {Kumar}, {Kanekar},
  \& {Bernardi}}]{2024MNRAS.528.5692K}
{Keller}, P.~M., {Thyagarajan}, N., {Kumar}, A., {Kanekar}, N., \& {Bernardi},
  G. 2024, \mnras, 528, 5692, \dodoi{10.1093/mnras/stae418}

\bibitem[{{Kellermann} {et~al.}(1989){Kellermann}, {Sramek}, {Schmidt},
  {Shaffer}, \& {Green}}]{1989AJ.....98.1195K}
{Kellermann}, K.~I., {Sramek}, R., {Schmidt}, M., {Shaffer}, D.~B., \& {Green},
  R. 1989, \aj, 98, 1195, \dodoi{10.1086/115207}

\bibitem[{{King}(2003)}]{2003ApJ...596L..27K}
{King}, A. 2003, \apjl, 596, L27, \dodoi{10.1086/379143}

\bibitem[{{Kocevski} {et~al.}(2023){Kocevski}, {Onoue}, {Inayoshi}, {Trump},
  {Arrabal Haro}, {Grazian}, {Dickinson}, {Finkelstein}, {Kartaltepe},
  {Hirschmann}, {Aird}, {Holwerda}, {Fujimoto}, {Juneau}, {Amor{\'\i}n},
  {Backhaus}, {Bagley}, {Barro}, {Bell}, {Bisigello}, {Calabr{\`o}}, {Cleri},
  {Cooper}, {Ding}, {Grogin}, {Ho}, {Hutchison}, {Inoue}, {Jiang}, {Jones},
  {Koekemoer}, {Li}, {Li}, {McGrath}, {Molina}, {Papovich},
  {P{\'e}rez-Gonz{\'a}lez}, {Pirzkal}, {Wilkins}, {Yang}, \&
  {Yung}}]{2023ApJ...954L...4K}
{Kocevski}, D.~D., {Onoue}, M., {Inayoshi}, K., {et~al.} 2023, \apjl, 954, L4,
  \dodoi{10.3847/2041-8213/ace5a0}

\bibitem[{{Kondapally} {et~al.}(2022){Kondapally}, {Best}, {Cochrane},
  {Sabater}, {Duncan}, {Hardcastle}, {Haskell}, {Mingo}, {R{\"o}ttgering},
  {Smith}, {Williams}, {Bonato}, {Calistro Rivera}, {Gao}, {Hale}, {Ma{\l}ek},
  {Miley}, {Prandoni}, \& {Wang}}]{2022MNRAS.513.3742K}
{Kondapally}, R., {Best}, P.~N., {Cochrane}, R.~K., {et~al.} 2022, \mnras, 513,
  3742, \dodoi{10.1093/mnras/stac1128}

\bibitem[{{Koopmans} {et~al.}(2015){Koopmans}, {Pritchard}, {Mellema},
  {Aguirre}, {Ahn}, {Barkana}, {van Bemmel}, {Bernardi}, {Bonaldi}, {Briggs},
  {de Bruyn}, {Chang}, {Chapman}, {Chen}, {Ciardi}, {Dayal}, {Ferrara},
  {Fialkov}, {Fiore}, {Ichiki}, {Illiev}, {Inoue}, {Jelic}, {Jones}, {Lazio},
  {Maio}, {Majumdar}, {Mack}, {Mesinger}, {Morales}, {Parsons}, {Pen},
  {Santos}, {Schneider}, {Semelin}, {de Souza}, {Subrahmanyan}, {Takeuchi},
  {Vedantham}, {Wagg}, {Webster}, {Wyithe}, {Datta}, \&
  {Trott}}]{2015aska.confE...1K}
{Koopmans}, L., {Pritchard}, J., {Mellema}, G., {et~al.} 2015, in Advancing
  Astrophysics with the Square Kilometre Array (AASKA14), 1,
  \dodoi{10.22323/1.215.0001}

\bibitem[{{Kormendy} \& {Ho}(2013)}]{2013ARA&A..51..511K}
{Kormendy}, J., \& {Ho}, L.~C. 2013, \araa, 51, 511,
  \dodoi{10.1146/annurev-astro-082708-101811}

\bibitem[{{Lambrides} {et~al.}(2024){Lambrides}, {Chiaberge}, {Long}, {Liu},
  {Akins}, {Ptak}, {Andika}, {Capetti}, {Casey}, {Champagne}, {Chworowsky},
  {Clarke}, {Cooper}, {Ding}, {Dong}, {Faisst}, {Forman}, {Franco}, {Gillman},
  {Gozaliasl}, {Hall}, {Harish}, {Hayward}, {Hirschmann}, {Hutchison},
  {Jahnke}, {Jin}, {Kartaltepe}, {Kleiner}, {Koekemoer}, {Kokorev}, {Manning},
  {Martin}, {McKinney}, {Norman}, {Nyland}, {Onoue}, {Robertson}, {Shuntov},
  {Silverman}, {Stiavelli}, {Trakhtenbrot}, {Vardoulaki}, {Zavala}, {Allen},
  {Ilbert}, {McCracken}, {Paquereau}, {Rhodes}, \&
  {Toft}}]{2024ApJ...961L..25L}
{Lambrides}, E., {Chiaberge}, M., {Long}, A.~S., {et~al.} 2024, \apjl, 961,
  L25, \dodoi{10.3847/2041-8213/ad11ee}

\bibitem[{{Larson} {et~al.}(2023){Larson}, {Finkelstein}, {Kocevski},
  {Hutchison}, {Trump}, {Arrabal Haro}, {Bromm}, {Cleri}, {Dickinson},
  {Fujimoto}, {Kartaltepe}, {Koekemoer}, {Papovich}, {Pirzkal}, {Tacchella},
  {Zavala}, {Bagley}, {Behroozi}, {Champagne}, {Cole}, {Jung}, {Morales},
  {Yang}, {Zhang}, {Zitrin}, {Amor{\'\i}n}, {Burgarella}, {Casey}, {Ch{\'a}vez
  Ortiz}, {Cox}, {Chworowsky}, {Fontana}, {Gawiser}, {Grazian}, {Grogin},
  {Harish}, {Hathi}, {Hirschmann}, {Holwerda}, {Juneau}, {Leung}, {Lucas},
  {McGrath}, {P{\'e}rez-Gonz{\'a}lez}, {Rigby}, {Seill{\'e}}, {Simons}, {de La
  Vega}, {Weiner}, {Wilkins}, {Yung}, \& {Ceers Team}}]{2023ApJ...953L..29L}
{Larson}, R.~L., {Finkelstein}, S.~L., {Kocevski}, D.~D., {et~al.} 2023, \apjl,
  953, L29, \dodoi{10.3847/2041-8213/ace619}

\bibitem[{{Li} {et~al.}(2023){Li}, {Inayoshi}, {Onoue}, \&
  {Toyouchi}}]{2023ApJ...950...85L}
{Li}, W., {Inayoshi}, K., {Onoue}, M., \& {Toyouchi}, D. 2023, \apj, 950, 85,
  \dodoi{10.3847/1538-4357/accbbe}

\bibitem[{{Lin} {et~al.}(2024){Lin}, {Wang}, {Fan}, {Cai}, {Champagne}, {Sun},
  {Volonteri}, {Yang}, {Hennawi}, {Ba{\~n}ados}, {Barth}, {Eilers}, {Farina},
  {Liu}, {Jin}, {Jun}, {Lupi}, {Kakiichi}, {Mazzucchelli}, {Onoue}, {Pan},
  {Pizzati}, {Rojas-Ruiz}, {Schindler}, {Trakhtenbrot}, {Shen}, {Trebitsch},
  {Zhuang}, {Endsley}, {Meyer}, {Li}, {Li}, {Pudoka}, {Tee}, {Wu}, \&
  {Zhang}}]{2024ApJ...974..147L}
{Lin}, X., {Wang}, F., {Fan}, X., {et~al.} 2024, \apj, 974, 147,
  \dodoi{10.3847/1538-4357/ad6565}

\bibitem[{{Liu} {et~al.}(2021){Liu}, {Wang}, {Momjian}, {Ba{\~n}ados},
  {Zeimann}, {Willott}, {Matsuoka}, {Omont}, {Shao}, {Li}, \&
  {Li}}]{2021ApJ...908..124L}
{Liu}, Y., {Wang}, R., {Momjian}, E., {et~al.} 2021, \apj, 908, 124,
  \dodoi{10.3847/1538-4357/abd3a8}

\bibitem[{{Lupi} {et~al.}(2023){Lupi}, {Quadri}, {Volonteri}, {Colpi}, \&
  {Regan}}]{2023arXiv231208422L}
{Lupi}, A., {Quadri}, G., {Volonteri}, M., {Colpi}, M., \& {Regan}, J.~A. 2023,
  arXiv e-prints, arXiv:2312.08422, \dodoi{10.48550/arXiv.2312.08422}

\bibitem[{{Lyu} {et~al.}(2024){Lyu}, {Alberts}, {Rieke}, {Shivaei},
  {P{\'e}rez-Gonz{\'a}lez}, {Sun}, {Hainline}, {Baum}, {Bonaventura}, {Bunker},
  {Egami}, {Eisenstein}, {Florian}, {Ji}, {Johnson}, {Morrison}, {Rieke},
  {Robertson}, {Rujopakarn}, {Tacchella}, {Scholtz}, \&
  {Willmer}}]{2024ApJ...966..229L}
{Lyu}, J., {Alberts}, S., {Rieke}, G.~H., {et~al.} 2024, \apj, 966, 229,
  \dodoi{10.3847/1538-4357/ad3643}

\bibitem[{{Mack} \& {Wyithe}(2012)}]{Mack2012}
{Mack}, K.~J., \& {Wyithe}, J. S.~B. 2012, \mnras, 425, 2988,
  \dodoi{10.1111/j.1365-2966.2012.21561.x}

\bibitem[{{Matsuoka} {et~al.}(2018){Matsuoka}, {Strauss}, {Kashikawa}, {Onoue},
  {Iwasawa}, {Tang}, {Lee}, {Imanishi}, {Nagao}, {Akiyama}, {Asami}, {Bosch},
  {Furusawa}, {Goto}, {Gunn}, {Harikane}, {Ikeda}, {Izumi}, {Kawaguchi},
  {Kato}, {Kikuta}, {Kohno}, {Komiyama}, {Lupton}, {Minezaki}, {Miyazaki},
  {Murayama}, {Niida}, {Nishizawa}, {Noboriguchi}, {Oguri}, {Ono}, {Ouchi},
  {Price}, {Sameshima}, {Schulze}, {Shirakata}, {Silverman}, {Sugiyama},
  {Tait}, {Takada}, {Takata}, {Tanaka}, {Toba}, {Utsumi}, {Wang}, \&
  {Yamashita}}]{2018ApJ...869..150M}
{Matsuoka}, Y., {Strauss}, M.~A., {Kashikawa}, N., {et~al.} 2018, \apj, 869,
  150, \dodoi{10.3847/1538-4357/aaee7a}

\bibitem[{{Matsuoka} {et~al.}(2023){Matsuoka}, {Onoue}, {Iwasawa}, {Strauss},
  {Kashikawa}, {Izumi}, {Nagao}, {Imanishi}, {Akiyama}, {Silverman}, {Asami},
  {Bosch}, {Furusawa}, {Goto}, {Gunn}, {Harikane}, {Ikeda}, {Inayoshi},
  {Ishimoto}, {Kawaguchi}, {Kikuta}, {Kohno}, {Komiyama}, {Lee}, {Lupton},
  {Minezaki}, {Miyazaki}, {Murayama}, {Nishizawa}, {Oguri}, {Ono}, {Oogi},
  {Ouchi}, {Price}, {Sameshima}, {Sugiyama}, {Tait}, {Takada}, {Takahashi},
  {Takata}, {Tanaka}, {Toba}, {Wang}, \& {Yamashita}}]{2023ApJ...949L..42M}
{Matsuoka}, Y., {Onoue}, M., {Iwasawa}, K., {et~al.} 2023, \apjl, 949, L42,
  \dodoi{10.3847/2041-8213/acd69f}

\bibitem[{{McConnell} \& {Ma}(2013)}]{2013ApJ...764..184M}
{McConnell}, N.~J., \& {Ma}, C.-P. 2013, \apj, 764, 184,
  \dodoi{10.1088/0004-637X/764/2/184}

\bibitem[{{McGreer} {et~al.}(2006){McGreer}, {Becker}, {Helfand}, \&
  {White}}]{2006ApJ...652..157M}
{McGreer}, I.~D., {Becker}, R.~H., {Helfand}, D.~J., \& {White}, R.~L. 2006,
  \apj, 652, 157, \dodoi{10.1086/507767}

\bibitem[{{Momjian} {et~al.}(2018){Momjian}, {Carilli}, {Ba{\~n}ados},
  {Walter}, \& {Venemans}}]{2018ApJ...861...86M}
{Momjian}, E., {Carilli}, C.~L., {Ba{\~n}ados}, E., {Walter}, F., \&
  {Venemans}, B.~P. 2018, \apj, 861, 86, \dodoi{10.3847/1538-4357/aac76f}

\bibitem[{{Momjian} {et~al.}(2008){Momjian}, {Carilli}, \&
  {McGreer}}]{2008AJ....136..344M}
{Momjian}, E., {Carilli}, C.~L., \& {McGreer}, I.~D. 2008, \aj, 136, 344,
  \dodoi{10.1088/0004-6256/136/1/344}

\bibitem[{{Momjian} {et~al.}(2014){Momjian}, {Carilli}, {Walter}, \&
  {Venemans}}]{2014AJ....147....6M}
{Momjian}, E., {Carilli}, C.~L., {Walter}, F., \& {Venemans}, B. 2014, \aj,
  147, 6, \dodoi{10.1088/0004-6256/147/1/6}

\bibitem[{{Monsalve} {et~al.}(2024){Monsalve}, {Altamirano}, {Bidula},
  {Bustos}, {Bye}, {Chiang}, {D{\'\i}az}, {Fern{\'a}ndez}, {Guo},
  {Hendricksen}, {Hornecker}, {Lucero}, {Mani}, {McGee}, {Mena}, {Pess{\^o}a},
  {Prabhakar}, {Restrepo}, {Sievers}, \& {Thyagarajan}}]{2024MNRAS.530.4125M}
{Monsalve}, R.~A., {Altamirano}, C., {Bidula}, V., {et~al.} 2024, \mnras, 530,
  4125, \dodoi{10.1093/mnras/stae1138}

\bibitem[{{Morales} \& {Wyithe}(2010)}]{2010ARA&A..48..127M}
{Morales}, M.~F., \& {Wyithe}, J. S.~B. 2010, \araa, 48, 127,
  \dodoi{10.1146/annurev-astro-081309-130936}

\bibitem[{{Morey} {et~al.}(2021){Morey}, {Eilers}, {Davies}, {Hennawi}, \&
  {Simcoe}}]{2021ApJ...921...88M}
{Morey}, K.~A., {Eilers}, A.-C., {Davies}, F.~B., {Hennawi}, J.~F., \&
  {Simcoe}, R.~A. 2021, \apj, 921, 88, \dodoi{10.3847/1538-4357/ac1c70}

\bibitem[{{Munshi} {et~al.}(2024){Munshi}, {Mertens}, {Koopmans}, {Offringa},
  {Semelin}, {Aubert}, {Barkana}, {Bracco}, {Brackenhoff}, {Cecconi},
  {Ceccotti}, {Corbel}, {Fialkov}, {Gehlot}, {Ghara}, {Girard},
  {Grie{\ss}meier}, {H{\"o}fer}, {Hothi}, {M{\'e}riot}, {Mevius}, {Ocvirk},
  {Shaw}, {Theureau}, {Yatawatta}, {Zarka}, \& {Zaroubi}}]{2024A&A...681A..62M}
{Munshi}, S., {Mertens}, F.~G., {Koopmans}, L.~V.~E., {et~al.} 2024, \aap, 681,
  A62, \dodoi{10.1051/0004-6361/202348329}

\bibitem[{Navros(2022)}]{Navros:2022qxa}
Navros, O. 2022, PhD thesis, Carnegie Mellon U. (main),
  \dodoi{10.1184/R1/19108616.v1}

\bibitem[{{Pacucci} {et~al.}(2015){Pacucci}, {Ferrara}, {Volonteri}, \&
  {Dubus}}]{2015MNRAS.454.3771P}
{Pacucci}, F., {Ferrara}, A., {Volonteri}, M., \& {Dubus}, G. 2015, \mnras,
  454, 3771, \dodoi{10.1093/mnras/stv2196}

\bibitem[{{Parsons} {et~al.}(2010){Parsons}, {Backer}, {Foster}, {Wright},
  {Bradley}, {Gugliucci}, {Parashare}, {Benoit}, {Aguirre}, {Jacobs},
  {Carilli}, {Herne}, {Lynch}, {Manley}, \& {Werthimer}}]{2010AJ....139.1468P}
{Parsons}, A.~R., {Backer}, D.~C., {Foster}, G.~S., {et~al.} 2010, \aj, 139,
  1468, \dodoi{10.1088/0004-6256/139/4/1468}

\bibitem[{{Patil} {et~al.}(2017){Patil}, {Yatawatta}, {Koopmans}, {de Bruyn},
  {Brentjens}, {Zaroubi}, {Asad}, {Hatef}, {Jeli{\'c}}, {Mevius}, {Offringa},
  {Pandey}, {Vedantham}, {Abdalla}, {Brouw}, {Chapman}, {Ciardi}, {Gehlot},
  {Ghosh}, {Harker}, {Iliev}, {Kakiichi}, {Majumdar}, {Mellema}, {Silva},
  {Schaye}, {Vrbanec}, \& {Wijnholds}}]{2017ApJ...838...65P}
{Patil}, A.~H., {Yatawatta}, S., {Koopmans}, L.~V.~E., {et~al.} 2017, \apj,
  838, 65, \dodoi{10.3847/1538-4357/aa63e7}

\bibitem[{{Philip} {et~al.}(2019){Philip}, {Abdurashidova}, {Chiang}, {Ghazi},
  {Gumba}, {Heilgendorff}, {J{\'a}uregui-Garc{\'\i}a}, {Malepe}, {Nunhokee},
  {Peterson}, {Sievers}, {Simes}, \& {Spann}}]{2019JAI.....850004P}
{Philip}, L., {Abdurashidova}, Z., {Chiang}, H.~C., {et~al.} 2019, JAI, 8,
  1950004, \dodoi{10.1142/S2251171719500041}

\bibitem[{{Planck Collaboration} {et~al.}(2020){Planck Collaboration},
  {Aghanim}, {Akrami}, {Ashdown}, {Aumont}, {Baccigalupi}, {Ballardini},
  {Banday}, {Barreiro}, {Bartolo}, {Basak}, {Battye}, {Benabed}, {Bernard},
  {Bersanelli}, {Bielewicz}, {Bock}, {Bond}, {Borrill}, {Bouchet}, {Boulanger},
  {Bucher}, {Burigana}, {Butler}, {Calabrese}, {Cardoso}, {Carron},
  {Challinor}, {Chiang}, {Chluba}, {Colombo}, {Combet}, {Contreras}, {Crill},
  {Cuttaia}, {de Bernardis}, {de Zotti}, {Delabrouille}, {Delouis}, {Di
  Valentino}, {Diego}, {Dor{\'e}}, {Douspis}, {Ducout}, {Dupac}, {Dusini},
  {Efstathiou}, {Elsner}, {En{\ss}lin}, {Eriksen}, {Fantaye}, {Farhang},
  {Fergusson}, {Fernandez-Cobos}, {Finelli}, {Forastieri}, {Frailis},
  {Fraisse}, {Franceschi}, {Frolov}, {Galeotta}, {Galli}, {Ganga},
  {G{\'e}nova-Santos}, {Gerbino}, {Ghosh}, {Gonz{\'a}lez-Nuevo}, {G{\'o}rski},
  {Gratton}, {Gruppuso}, {Gudmundsson}, {Hamann}, {Handley}, {Hansen},
  {Herranz}, {Hildebrandt}, {Hivon}, {Huang}, {Jaffe}, {Jones}, {Karakci},
  {Keih{\"a}nen}, {Keskitalo}, {Kiiveri}, {Kim}, {Kisner}, {Knox},
  {Krachmalnicoff}, {Kunz}, {Kurki-Suonio}, {Lagache}, {Lamarre}, {Lasenby},
  {Lattanzi}, {Lawrence}, {Le Jeune}, {Lemos}, {Lesgourgues}, {Levrier},
  {Lewis}, {Liguori}, {Lilje}, {Lilley}, {Lindholm}, {L{\'o}pez-Caniego},
  {Lubin}, {Ma}, {Mac{\'\i}as-P{\'e}rez}, {Maggio}, {Maino}, {Mandolesi},
  {Mangilli}, {Marcos-Caballero}, {Maris}, {Martin}, {Martinelli},
  {Mart{\'\i}nez-Gonz{\'a}lez}, {Matarrese}, {Mauri}, {McEwen}, {Meinhold},
  {Melchiorri}, {Mennella}, {Migliaccio}, {Millea}, {Mitra},
  {Miville-Desch{\^e}nes}, {Molinari}, {Montier}, {Morgante}, {Moss}, {Natoli},
  {N{\o}rgaard-Nielsen}, {Pagano}, {Paoletti}, {Partridge}, {Patanchon},
  {Peiris}, {Perrotta}, {Pettorino}, {Piacentini}, {Polastri}, {Polenta},
  {Puget}, {Rachen}, {Reinecke}, {Remazeilles}, {Renzi}, {Rocha}, {Rosset},
  {Roudier}, {Rubi{\~n}o-Mart{\'\i}n}, {Ruiz-Granados}, {Salvati}, {Sandri},
  {Savelainen}, {Scott}, {Shellard}, {Sirignano}, {Sirri}, {Spencer},
  {Sunyaev}, {Suur-Uski}, {Tauber}, {Tavagnacco}, {Tenti}, {Toffolatti},
  {Tomasi}, {Trombetti}, {Valenziano}, {Valiviita}, {Van Tent}, {Vibert},
  {Vielva}, {Villa}, {Vittorio}, {Wandelt}, {Wehus}, {White}, {White},
  {Zacchei}, \& {Zonca}}]{2020A&A...641A...6P}
{Planck Collaboration}, {Aghanim}, N., {Akrami}, Y., {et~al.} 2020, \aap, 641,
  A6, \dodoi{10.1051/0004-6361/201833910}

\bibitem[{{Price} {et~al.}(2018){Price}, {Greenhill}, {Fialkov}, {Bernardi},
  {Garsden}, {Barsdell}, {Kocz}, {Anderson}, {Bourke}, {Craig}, {Dexter},
  {Dowell}, {Eastwood}, {Eftekhari}, {Ellingson}, {Hallinan}, {Hartman},
  {Kimberk}, {Lazio}, {Leiker}, {MacMahon}, {Monroe}, {Schinzel}, {Taylor},
  {Tong}, {Werthimer}, \& {Woody}}]{2018MNRAS.478.4193P}
{Price}, D.~C., {Greenhill}, L.~J., {Fialkov}, A., {et~al.} 2018, \mnras, 478,
  4193, \dodoi{10.1093/mnras/sty1244}

\bibitem[{{Pritchard} \& {Loeb}(2012)}]{2012RPPh...75h6901P}
{Pritchard}, J.~R., \& {Loeb}, A. 2012, RPPh, 75, 086901,
  \dodoi{10.1088/0034-4885/75/8/086901}

\bibitem[{{Runnoe} {et~al.}(2012){Runnoe}, {Brotherton}, \&
  {Shang}}]{2012MNRAS.422..478R}
{Runnoe}, J.~C., {Brotherton}, M.~S., \& {Shang}, Z. 2012, \mnras, 422, 478,
  \dodoi{10.1111/j.1365-2966.2012.20620.x}

\bibitem[{{Sabater} {et~al.}(2021){Sabater}, {Best}, {Tasse}, {Hardcastle},
  {Shimwell}, {Nisbet}, {Jelic}, {Callingham}, {R{\"o}ttgering}, {Bonato},
  {Bondi}, {Ciardi}, {Cochrane}, {Jarvis}, {Kondapally}, {Koopmans},
  {O'Sullivan}, {Prandoni}, {Schwarz}, {Smith}, {Wang}, {Williams}, \&
  {Zaroubi}}]{2021A&A...648A...2S}
{Sabater}, J., {Best}, P.~N., {Tasse}, C., {et~al.} 2021, \aap, 648, A2,
  \dodoi{10.1051/0004-6361/202038828}

\bibitem[{{Sathyanarayana Rao} {et~al.}(2023){Sathyanarayana Rao}, {Singh},
  {K.~S.}, {B.~S.}, {Sathish}, {Somashekar}, {Agaram}, {Kavitha},
  {Vishwapriya}, {Anand}, {Udaya Shankar}, \& {Seetha}}]{2023ExA....56..741S}
{Sathyanarayana Rao}, M., {Singh}, S., {K.~S.}, S., {et~al.} 2023, ExA, 56,
  741, \dodoi{10.1007/s10686-023-09909-5}

\bibitem[{{Schmidt}(1968)}]{1968ApJ...151..393S}
{Schmidt}, M. 1968, \apj, 151, 393, \dodoi{10.1086/149446}

\bibitem[{{Shang} {et~al.}(2011){Shang}, {Brotherton}, {Wills}, {Wills},
  {Cales}, {Dale}, {Green}, {Runnoe}, {Nemmen}, {Gallagher}, {Ganguly},
  {Hines}, {Kelly}, {Kriss}, {Li}, {Tang}, \& {Xie}}]{2011ApJS..196....2S}
{Shang}, Z., {Brotherton}, M.~S., {Wills}, B.~J., {et~al.} 2011, \apjs, 196, 2,
  \dodoi{10.1088/0067-0049/196/1/2}

\bibitem[{{Shao} {et~al.}(2023){Shao}, {Xu}, {Wang}, {Yang}, {Li}, {Zhang}, \&
  {Chen}}]{2023NatAs...7.1116S}
{Shao}, Y., {Xu}, Y., {Wang}, Y., {et~al.} 2023, NatAs, 7, 1116,
  \dodoi{10.1038/s41550-023-02024-7}

\bibitem[{{Sheth} \& {Tormen}(2002)}]{2002MNRAS.329...61S}
{Sheth}, R.~K., \& {Tormen}, G. 2002, \mnras, 329, 61,
  \dodoi{10.1046/j.1365-8711.2002.04950.x}

\bibitem[{{Shimabukuro} {et~al.}(2014){Shimabukuro}, {Ichiki}, {Inoue}, \&
  {Yokoyama}}]{2014PhRvD..90h3003S}
{Shimabukuro}, H., {Ichiki}, K., {Inoue}, S., \& {Yokoyama}, S. 2014, \prd, 90,
  083003, \dodoi{10.1103/PhysRevD.90.083003}

\bibitem[{{Shimabukuro} {et~al.}(2020){Shimabukuro}, {Ichiki}, \&
  {Kadota}}]{2020PhRvD.101d3516S}
{Shimabukuro}, H., {Ichiki}, K., \& {Kadota}, K. 2020, \prd, 101, 043516,
  \dodoi{10.1103/PhysRevD.101.043516}

\bibitem[{{Shimasaku} \& {Izumi}(2019)}]{2019ApJ...872L..29S}
{Shimasaku}, K., \& {Izumi}, T. 2019, \apjl, 872, L29,
  \dodoi{10.3847/2041-8213/ab053f}

\bibitem[{{Silk} \& {Rees}(1998)}]{1998A&A...331L...1S}
{Silk}, J., \& {Rees}, M.~J. 1998, \aap, 331, L1,
  \dodoi{10.48550/arXiv.astro-ph/9801013}

\bibitem[{{Singh} {et~al.}(2018){Singh}, {Subrahmanyan}, {Udaya Shankar},
  {Sathyanarayana Rao}, {Fialkov}, {Cohen}, {Barkana}, {Girish}, {Raghunathan},
  {Somashekar}, \& {Srivani}}]{2018ApJ...858...54S}
{Singh}, S., {Subrahmanyan}, R., {Udaya Shankar}, N., {et~al.} 2018, \apj, 858,
  54, \dodoi{10.3847/1538-4357/aabae1}

\bibitem[{{Spergel} {et~al.}(2015){Spergel}, {Gehrels}, {Baltay}, {Bennett},
  {Breckinridge}, {Donahue}, {Dressler}, {Gaudi}, {Greene}, {Guyon}, {Hirata},
  {Kalirai}, {Kasdin}, {Macintosh}, {Moos}, {Perlmutter}, {Postman},
  {Rauscher}, {Rhodes}, {Wang}, {Weinberg}, {Benford}, {Hudson}, {Jeong},
  {Mellier}, {Traub}, {Yamada}, {Capak}, {Colbert}, {Masters}, {Penny},
  {Savransky}, {Stern}, {Zimmerman}, {Barry}, {Bartusek}, {Carpenter}, {Cheng},
  {Content}, {Dekens}, {Demers}, {Grady}, {Jackson}, {Kuan}, {Kruk}, {Melton},
  {Nemati}, {Parvin}, {Poberezhskiy}, {Peddie}, {Ruffa}, {Wallace}, {Whipple},
  {Wollack}, \& {Zhao}}]{2015arXiv150303757S}
{Spergel}, D., {Gehrels}, N., {Baltay}, C., {et~al.} 2015, arXiv e-prints,
  arXiv:1503.03757, \dodoi{10.48550/arXiv.1503.03757}

\bibitem[{{Square Kilometre Array Cosmology Science Working Group}
  {et~al.}(2020){Square Kilometre Array Cosmology Science Working Group},
  {Bacon}, {Battye}, {Bull}, {Camera}, {Ferreira}, {Harrison}, {Parkinson},
  {Pourtsidou}, {Santos}, {Wolz}, {Abdalla}, {Akrami}, {Alonso},
  {Andrianomena}, {Ballardini}, {Bernal}, {Bertacca}, {Bengaly}, {Bonaldi},
  {Bonvin}, {Brown}, {Chapman}, {Chen}, {Chen}, {Cunnington}, {Davis},
  {Dickinson}, {Fonseca}, {Grainge}, {Harper}, {Jarvis}, {Maartens}, {Maddox},
  {Padmanabhan}, {Pritchard}, {Raccanelli}, {Rivi}, {Roychowdhury},
  {Sahl{\'e}n}, {Schwarz}, {Siewert}, {Viel}, {Villaescusa-Navarro}, {Xu},
  {Yamauchi}, \& {Zuntz}}]{2020PASA...37....7S}
{Square Kilometre Array Cosmology Science Working Group}, {Bacon}, D.~J.,
  {Battye}, R.~A., {et~al.} 2020, \pasa, 37, e007, \dodoi{10.1017/pasa.2019.51}

\bibitem[{{Stern} {et~al.}(2000){Stern}, {Djorgovski}, {Perley}, {de Carvalho},
  \& {Wall}}]{2000AJ....119.1526S}
{Stern}, D., {Djorgovski}, S.~G., {Perley}, R.~A., {de Carvalho}, R.~R., \&
  {Wall}, J.~V. 2000, \aj, 119, 1526, \dodoi{10.1086/301316}

\bibitem[{{Strittmatter} {et~al.}(1980){Strittmatter}, {Hill}, {Pauliny-Toth},
  {Steppe}, \& {Witzel}}]{1980A&A....88L..12S}
{Strittmatter}, P.~A., {Hill}, P., {Pauliny-Toth}, I.~I.~K., {Steppe}, H., \&
  {Witzel}, A. 1980, \aap, 88, L12

\bibitem[{{Sun} {et~al.}(2024){Sun}, {Shao}, {Li}, {Xu}, \&
  {Zhang}}]{2024arXiv240714298S}
{Sun}, T.-Y., {Shao}, Y., {Li}, Y., {Xu}, Y., \& {Zhang}, X. 2024, arXiv
  e-prints, arXiv:2407.14298, \dodoi{10.48550/arXiv.2407.14298}

\bibitem[{{Thompson} {et~al.}(2017){Thompson}, {Moran}, \&
  {Swenson}}]{2017isra.book.....T}
{Thompson}, A.~R., {Moran}, J.~M., \& {Swenson}, George~W., J. 2017,
  {Interferometry and Synthesis in Radio Astronomy, 3rd Edition},
  \dodoi{10.1007/978-3-319-44431-4}

\bibitem[{{Thyagarajan}(2020)}]{Thyagarajan2020}
{Thyagarajan}, N. 2020, \apj, 899, 16, \dodoi{10.3847/1538-4357/ab9e6d}

\bibitem[{{Tingay} {et~al.}(2013){Tingay}, {Goeke}, {Bowman}, {Emrich}, {Ord},
  {Mitchell}, {Morales}, {Booler}, {Crosse}, {Wayth}, {Lonsdale}, {Tremblay},
  {Pallot}, {Colegate}, {Wicenec}, {Kudryavtseva}, {Arcus}, {Barnes},
  {Bernardi}, {Briggs}, {Burns}, {Bunton}, {Cappallo}, {Corey}, {Deshpande},
  {Desouza}, {Gaensler}, {Greenhill}, {Hall}, {Hazelton}, {Herne}, {Hewitt},
  {Johnston-Hollitt}, {Kaplan}, {Kasper}, {Kincaid}, {Koenig}, {Kratzenberg},
  {Lynch}, {Mckinley}, {Mcwhirter}, {Morgan}, {Oberoi}, {Pathikulangara},
  {Prabu}, {Remillard}, {Rogers}, {Roshi}, {Salah}, {Sault}, {Udaya-Shankar},
  {Schlagenhaufer}, {Srivani}, {Stevens}, {Subrahmanyan}, {Waterson},
  {Webster}, {Whitney}, {Williams}, {Williams}, \&
  {Wyithe}}]{2013PASA...30....7T}
{Tingay}, S.~J., {Goeke}, R., {Bowman}, J.~D., {et~al.} 2013, \pasa, 30, e007,
  \dodoi{10.1017/pasa.2012.007}

\bibitem[{{{\"U}bler} {et~al.}(2023){{\"U}bler}, {Maiolino}, {Curtis-Lake},
  {P{\'e}rez-Gonz{\'a}lez}, {Curti}, {Perna}, {Arribas}, {Charlot}, {Marshall},
  {D'Eugenio}, {Scholtz}, {Bunker}, {Carniani}, {Ferruit}, {Jakobsen}, {Rix},
  {Rodr{\'\i}guez Del Pino}, {Willott}, {Boeker}, {Cresci}, {Jones}, {Kumari},
  \& {Rawle}}]{2023A&A...677A.145U}
{{\"U}bler}, H., {Maiolino}, R., {Curtis-Lake}, E., {et~al.} 2023, \aap, 677,
  A145, \dodoi{10.1051/0004-6361/202346137}

\bibitem[{{van Haarlem} {et~al.}(2013){van Haarlem}, {Wise}, {Gunst}, {Heald},
  {McKean}, {Hessels}, {de Bruyn}, {Nijboer}, {Swinbank}, {Fallows},
  {Brentjens}, {Nelles}, {Beck}, {Falcke}, {Fender}, {H{\"o}randel},
  {Koopmans}, {Mann}, {Miley}, {R{\"o}ttgering}, {Stappers}, {Wijers},
  {Zaroubi}, {van den Akker}, {Alexov}, {Anderson}, {Anderson}, {van Ardenne},
  {Arts}, {Asgekar}, {Avruch}, {Batejat}, {B{\"a}hren}, {Bell}, {Bell}, {van
  Bemmel}, {Bennema}, {Bentum}, {Bernardi}, {Best}, {B{\^\i}rzan}, {Bonafede},
  {Boonstra}, {Braun}, {Bregman}, {Breitling}, {van de Brink}, {Broderick},
  {Broekema}, {Brouw}, {Br{\"u}ggen}, {Butcher}, {van Cappellen}, {Ciardi},
  {Coenen}, {Conway}, {Coolen}, {Corstanje}, {Damstra}, {Davies}, {Deller},
  {Dettmar}, {van Diepen}, {Dijkstra}, {Donker}, {Doorduin}, {Dromer}, {Drost},
  {van Duin}, {Eisl{\"o}ffel}, {van Enst}, {Ferrari}, {Frieswijk}, {Gankema},
  {Garrett}, {de Gasperin}, {Gerbers}, {de Geus}, {Grie{\ss}meier}, {Grit},
  {Gruppen}, {Hamaker}, {Hassall}, {Hoeft}, {Holties}, {Horneffer}, {van der
  Horst}, {van Houwelingen}, {Huijgen}, {Iacobelli}, {Intema}, {Jackson},
  {Jelic}, {de Jong}, {Juette}, {Kant}, {Karastergiou}, {Koers}, {Kollen},
  {Kondratiev}, {Kooistra}, {Koopman}, {Koster}, {Kuniyoshi}, {Kramer},
  {Kuper}, {Lambropoulos}, {Law}, {van Leeuwen}, {Lemaitre}, {Loose}, {Maat},
  {Macario}, {Markoff}, {Masters}, {McFadden}, {McKay-Bukowski}, {Meijering},
  {Meulman}, {Mevius}, {Middelberg}, {Millenaar}, {Miller-Jones}, {Mohan},
  {Mol}, {Morawietz}, {Morganti}, {Mulcahy}, {Mulder}, {Munk}, {Nieuwenhuis},
  {van Nieuwpoort}, {Noordam}, {Norden}, {Noutsos}, {Offringa}, {Olofsson},
  {Omar}, {Orr{\'u}}, {Overeem}, {Paas}, {Pandey-Pommier}, {Pandey}, {Pizzo},
  {Polatidis}, {Rafferty}, {Rawlings}, {Reich}, {de Reijer}, {Reitsma},
  {Renting}, {Riemers}, {Rol}, {Romein}, {Roosjen}, {Ruiter}, {Scaife}, {van
  der Schaaf}, {Scheers}, {Schellart}, {Schoenmakers}, {Schoonderbeek},
  {Serylak}, {Shulevski}, {Sluman}, {Smirnov}, {Sobey}, {Spreeuw}, {Steinmetz},
  {Sterks}, {Stiepel}, {Stuurwold}, {Tagger}, {Tang}, {Tasse}, {Thomas},
  {Thoudam}, {Toribio}, {van der Tol}, {Usov}, {van Veelen}, {van der Veen},
  {ter Veen}, {Verbiest}, {Vermeulen}, {Vermaas}, {Vocks}, {Vogt}, {de Vos},
  {van der Wal}, {van Weeren}, {Weggemans}, {Weltevrede}, {White}, {Wijnholds},
  {Wilhelmsson}, {Wucknitz}, {Yatawatta}, {Zarka}, {Zensus}, \& {van
  Zwieten}}]{2013A&A...556A...2V}
{van Haarlem}, M.~P., {Wise}, M.~W., {Gunst}, A.~W., {et~al.} 2013, \aap, 556,
  A2, \dodoi{10.1051/0004-6361/201220873}

\bibitem[{{{\v{S}}oltinsk{\'y}} {et~al.}(2023){{\v{S}}oltinsk{\'y}}, {Bolton},
  {Molaro}, {Hatch}, {Haehnelt}, {Keating}, {Kulkarni}, \&
  {Puchwein}}]{2023MNRAS.519.3027S}
{{\v{S}}oltinsk{\'y}}, T., {Bolton}, J.~S., {Molaro}, M., {et~al.} 2023,
  \mnras, 519, 3027, \dodoi{10.1093/mnras/stac3710}

\bibitem[{{{\v{S}}oltinsk{\'y}} {et~al.}(2021){{\v{S}}oltinsk{\'y}}, {Bolton},
  {Hatch}, {Haehnelt}, {Keating}, {Kulkarni}, {Puchwein}, {Chardin}, \&
  {Aubert}}]{2021MNRAS.506.5818S}
{{\v{S}}oltinsk{\'y}}, T., {Bolton}, J.~S., {Hatch}, N., {et~al.} 2021, \mnras,
  506, 5818, \dodoi{10.1093/mnras/stab1830}

\bibitem[{{Wang} {et~al.}(2021){Wang}, {Yang}, {Fan}, {Hennawi}, {Barth},
  {Banados}, {Bian}, {Boutsia}, {Connor}, {Davies}, {Decarli}, {Eilers},
  {Farina}, {Green}, {Jiang}, {Li}, {Mazzucchelli}, {Nanni}, {Schindler},
  {Venemans}, {Walter}, {Wu}, \& {Yue}}]{2021ApJ...907L...1W}
{Wang}, F., {Yang}, J., {Fan}, X., {et~al.} 2021, \apjl, 907, L1,
  \dodoi{10.3847/2041-8213/abd8c6}

\bibitem[{{Wang} {et~al.}(2007){Wang}, {Carilli}, {Beelen}, {Bertoldi}, {Fan},
  {Walter}, {Menten}, {Omont}, {Cox}, {Strauss}, \&
  {Jiang}}]{2007AJ....134..617W}
{Wang}, R., {Carilli}, C.~L., {Beelen}, A., {et~al.} 2007, \aj, 134, 617,
  \dodoi{10.1086/518867}

\bibitem[{{White} {et~al.}(2007){White}, {Helfand}, {Becker}, {Glikman}, \& {de
  Vries}}]{2007ApJ...654...99W}
{White}, R.~L., {Helfand}, D.~J., {Becker}, R.~H., {Glikman}, E., \& {de
  Vries}, W. 2007, \apj, 654, 99, \dodoi{10.1086/507700}

\bibitem[{{Willott} {et~al.}(2010){Willott}, {Delorme}, {Reyl{\'e}}, {Albert},
  {Bergeron}, {Crampton}, {Delfosse}, {Forveille}, {Hutchings}, {McLure},
  {Omont}, \& {Schade}}]{2010AJ....139..906W}
{Willott}, C.~J., {Delorme}, P., {Reyl{\'e}}, C., {et~al.} 2010, \aj, 139, 906,
  \dodoi{10.1088/0004-6256/139/3/906}

\bibitem[{{Wyithe} \& {Loeb}(2003)}]{2003ApJ...595..614W}
{Wyithe}, J. S.~B., \& {Loeb}, A. 2003, \apj, 595, 614, \dodoi{10.1086/377475}

\bibitem[{{Xiao} {et~al.}(2022){Xiao}, {Zhu}, {Fu}, {Zhang}, \&
  {Fan}}]{2022PASJ...74..239X}
{Xiao}, H., {Zhu}, J., {Fu}, L., {Zhang}, S., \& {Fan}, J. 2022, \pasj, 74,
  239, \dodoi{10.1093/pasj/psab121}

\bibitem[{{Xu} {et~al.}(2009){Xu}, {Chen}, {Fan}, {Trac}, \&
  {Cen}}]{2009ApJ...704.1396X}
{Xu}, Y., {Chen}, X., {Fan}, Z., {Trac}, H., \& {Cen}, R. 2009, \apj, 704,
  1396, \dodoi{10.1088/0004-637X/704/2/1396}

\bibitem[{{Xu} {et~al.}(2011){Xu}, {Ferrara}, \& {Chen}}]{2011MNRAS.410.2025X}
{Xu}, Y., {Ferrara}, A., \& {Chen}, X. 2011, \mnras, 410, 2025,
  \dodoi{10.1111/j.1365-2966.2010.17579.x}

\bibitem[{{Xu} {et~al.}(2010){Xu}, {Ferrara}, {Kitaura}, \&
  {Chen}}]{2010SCPMA..53.1124X}
{Xu}, Y., {Ferrara}, A., {Kitaura}, F.~S., \& {Chen}, X. 2010, SCPMA, 53, 1124,
  \dodoi{10.1007/s11433-010-3208-x}

\bibitem[{{Xu} \& {Zhang}(2020)}]{2020SCPMA..6370431X}
{Xu}, Y., \& {Zhang}, X. 2020, SCPMA, 63, 270431,
  \dodoi{10.1007/s11433-020-1544-3}

\bibitem[{{Yang} {et~al.}(2021){Yang}, {Wang}, {Fan}, {Barth}, {Hennawi},
  {Nanni}, {Bian}, {Davies}, {Farina}, {Schindler}, {Ba{\~n}ados}, {Decarli},
  {Eilers}, {Green}, {Guo}, {Jiang}, {Li}, {Venemans}, {Walter}, {Wu}, \&
  {Yue}}]{2021ApJ...923..262Y}
{Yang}, J., {Wang}, F., {Fan}, X., {et~al.} 2021, \apj, 923, 262,
  \dodoi{10.3847/1538-4357/ac2b32}

\bibitem[{{Yang} {et~al.}(2023){Yang}, {Fan}, {Gupta}, {Myers},
  {Palanque-Delabrouille}, {Wang}, {Y{\`e}che}, {Aguilar}, {Ahlen},
  {Alexander}, {Brooks}, {Dawson}, {de la Macorra}, {Dey}, {Dhungana},
  {Fanning}, {Font-Ribera}, {Gontcho}, {Guy}, {Honscheid}, {Juneau}, {Kisner},
  {Kremin}, {Le Guillou}, {Levi}, {Magneville}, {Martini}, {Meisner}, {Miquel},
  {Moustakas}, {Nie}, {Percival}, {Poppett}, {Prada}, {Schlafly}, {Tarl{\'e}},
  {Vargas Magana}, {Weaver}, {Wechsler}, {Zhou}, {Zhou}, \&
  {Zou}}]{2023ApJS..269...27Y}
{Yang}, J., {Fan}, X., {Gupta}, A., {et~al.} 2023, \apjs, 269, 27,
  \dodoi{10.3847/1538-4365/acf99b}

\bibitem[{{Zubovas} \& {King}(2021)}]{2021MNRAS.501.4289Z}
{Zubovas}, K., \& {King}, A. 2021, \mnras, 501, 4289,
  \dodoi{10.1093/mnras/stab004}

\end{thebibliography}
\bibliographystyle{aasjournal}

\end{document}